\documentclass[iop]{emulateapj}
\usepackage{apjfonts}
\journalinfo{The Astrophysical Journal, 2013, in press}

\shorttitle{From the Magnetic Carpet to the Heliosphere}
\shortauthors{Cranmer, van Ballegooijen, \& Woolsey}

\begin{document}

\title{Connecting the Sun's High-Resolution Magnetic Carpet to
the Turbulent Heliosphere}

\author{Steven R. Cranmer,
Adriaan A. van Ballegooijen,
and Lauren N. Woolsey}

\affil{Harvard-Smithsonian Center for Astrophysics,
60 Garden Street, Cambridge, MA 02138, USA} 

\begin{abstract}
The solar wind is connected to the Sun's atmosphere by
flux tubes that are rooted in an ever-changing pattern of positive
and negative magnetic polarities on the surface.
Observations indicate that the magnetic field is filamentary
and intermittent across a wide range of spatial scales.
However, we do not know to what extent the complex flux tube
topology seen near the Sun survives as the wind expands into
interplanetary space.
In order to study the possible long-distance connections between
the corona and the heliosphere, we developed new models of
turbulence-driven solar wind acceleration along empirically
constrained field lines.
We used a potential-field model of the Quiet Sun 
to trace field lines into the ecliptic plane with unprecedented
spatial resolution at their footpoints.
For each flux tube, a one-dimensional model was created with an
existing wave/turbulence code that solves equations of mass,
momentum, and energy conservation from the photosphere to 4 AU.
To take account of stream-stream interactions between flux tubes,
we used those models as inner boundary conditions for a
time-steady MHD description of radial and longitudinal
structure in the ecliptic.
Corotating stream interactions smear out much of the
smallest-scale variability, making it difficult to see how
individual flux tubes on granular or supergranular scales can
survive out to 1 AU.
However, our models help clarify the level of ``background''
variability with which waves and turbulent eddies should be
expected to interact.
Also, the modeled fluctuations in magnetic field magnitude were
seen to match measured power spectra quite well.
\end{abstract}


\section{Introduction}
\label{sec:intro}

The Sun is known to vary on scales that span at least twenty orders
of magnitude in time---from milliseconds \citep{Ba98} to
gigayears \citep{Gu07}.
The corresponding spatial variability may not have such a huge dynamic
range, but even the best observations of the solar disk---now reaching
below $10^{-4}$ times the solar radius ($R_{\odot}$)---have
not yet fully resolved the existing magnetic structures.
There is also complex variability in the particles and electromagnetic
fields measured far from the Sun in interplanetary space
\citep{BC05,Pt10}.
It is a major goal of solar and heliospheric physics to better
understand how the changing conditions on the solar surface give
rise to variations at greater distances.

It has been suspected for many years that some trace of the
filamentary flux-tube structure of the solar corona---e.g., polar
plumes, jets, streamer cusps---may survive into interplanetary space
\citep{MN66,Th90,Ng95,Re99,Ya02,Dc05,Gc06,Bo08,Bo10}.
However, since heliospheric turbulence appears to be in a
reasonably developed state, it is unclear how pristine solar
flux tubes can avoid being completely dispersed or broken up
by the chaotic turbulent eddies \citep{Mu91,Mt95}.
Some simulations of magnetohydrodynamic (MHD) turbulence have
shown the ability to preserve input signals at single frequencies
\citep[e.g.,][]{Gh09}, so it is worthwhile to explore to what
extent the ``signals'' of solar flux tubes may survive as well.

The generation of small-scale structures in the Sun's magnetic field
depends to some degree on the physical processes that heat the
corona and accelerate the solar wind.
Although these problems still have no universally accepted solutions,
there are two broad types of explanation being debated.
First, there are models that rely on waves and turbulent motions
to propagate up the open-field flux tubes and dissipate to generate
the required thermal energy \citep{Mt99,CvB07,Ch11,Vd12a}.
Second, there are models that assume the wind's mass and energy
is injected from closed-field regions via magnetic reconnection
\citep{Fi99,Mo11,An11}.
Regardless of whether the dominant coronal fluctuations are
wave-like or reconnection-driven, they appear to be generated
at small spatial scales in the lower atmosphere and are magnified
and ``stretched'' as they evolve outward into the heliosphere.
Their impact on the solar wind's energy budget depends on the
multi-scale topological structure of the Sun's magnetic field.

In this paper, we aim to improve our understanding of the origin
of the turbulent solar wind by studying the complex structure of
open coronal flux tubes.
We will use high-resolution multidimensional models of solar
wind acceleration to begin addressing several of the following
questions.
For example, how much spatial resolution is really required when
modeling the field lines that connect the solar surface and the
solar wind?
Are there specific in~situ MHD fluctuations that can be
attributed to the survival of coronal flux tubes?
Does the production of corotating interaction regions (CIRs)
act as a net source for turbulence \citep{Za96} or is it
mainly a way of smearing out longitudinal gradients and thus
reducing the variability \citep[e.g.,][]{SW76,Go96,Ri07}?
Lastly, does the well-known empirical correlation between wind
speed and magnetic flux tube expansion \citep{WS90} remain
valid for topologically complex bundles of field lines rooted
in a mixed-polarity ``magnetic carpet?''

In Section~\ref{sec:solis} of this paper, we present a
high-resolution potential extrapolation of the magnetic field
from a quiescent source region of slow solar wind.
We also summarize the radial and longitudinal properties of
the open flux tubes in this model and discuss the solar wind
speeds that would be expected at 1~AU on the basis of existing
empirical correlations.
Section \ref{sec:nonpot} describes a set of non-potential
enhancements that we make to the field lines in order to
more accurately simulate solar wind acceleration regions.
Section \ref{sec:zephyr} summarizes the results of running these
flux tubes through the turbulence-driven model of coronal heating
and solar wind acceleration of \citet{CvB07}.
In Section \ref{sec:cir} we describe how we use the plasma
parameters from the one-dimensional solar wind models as inputs
to a two-dimensional description of CIR formation in the ecliptic
plane.
We also discuss the statistical properties of the modeled MHD
fluctuations in Section \ref{sec:flucts}.
Lastly, Section \ref{sec:conc} concludes this paper with a summary
of the major results, a discussion of some of the wider implications
of this work, and suggestions for future improvements.

\section{High-Resolution Quiet-Sun Magnetic Field}
\label{sec:solis}

Our goal is to evaluate the importance of resolving small-scale
magnetic features when tracing flux tubes connected to the
low-latitude slow solar wind.
Thus, we chose a time period during which the footpoints linked to
the ecliptic plane were rooted in regions of Quiet Sun (QS)
away from both unipolar coronal holes and strong-field active
regions.
Although QS regions are typically associated with closed loop-like
magnetic fields, it is likely that some fraction of quiescent
slow wind is associated with them as well \citep[e.g.,][]{He07,Ti11}.
Magnetically balanced QS regions, with only a small fraction of
their magnetic flux open to the heliosphere, also exhibit large
superradial expansion factors that are associated with slow wind
speeds \citep{WS90}.

We used magnetic field measurements made by the Vector
SpectroMagnetograph (VSM) instrument of the Synoptic Optical
Long-term Investigations of the Sun (SOLIS) facility
\citep{Kc03,Hn09}.
SOLIS magnetograms may not have the highest spatial resolution in
comparison to other available data sets, but its sensitivity to
weak fields made it the optimal choice for mapping accurate
footpoints of open QS flux tubes.
The primary measurement was a full-disk longitudinal magnetogram
taken in the \ion{Fe}{1} 6301.5 {\AA} line at time
$t_{0} =$~16:30 UT on 2003 September 4.
For global context, the central portions of this magnetogram were
embedded in a lower-resolution synoptic magnetogram for
Carrington Rotation (CR) 2007.
At time $t_{0}$, the central meridian longitude during this CR
was $\phi_{0} = {286.6\arcdeg}$.

The magnetic field was extrapolated from the photosphere to the corona
using the standard Potential Field Source Surface (PFSS) method, which
assumes the corona is current-free out to a spherical ``source surface''
above which the field is radial \citep{Sh69,AN69}.
The source surface radius was chosen as $r = 2.46 \, R_{\odot}$.
Details about the numerical method used to construct the PFSS model
are provided in Appendix \ref{appen:pfss}.
The PFSS method has been shown to create a relatively good mapping
between the Sun and the heliosphere \citep{AP00,Lu02,WS06},
although full MHD simulations of course take better account of
gas pressure effects and stream-stream interactions
(see Section \ref{sec:cir}).

We traced a set of ``open'' field lines down from the source surface
to the photospheric lower boundary.
The locus of starting points at $r = 2.46 \, R_{\odot}$ was
coincident with the ecliptic plane.
The longitudinal separation in the grid of starting points was
chosen to be equivalent to 1 minute of solar rotation time.
The total longitudinal extent of this region was {80\arcdeg}, which
placed the footpoints squarely inside the disk-center regions of
the high-resolution magnetogram obtained on 2003 September 4.
Using a Carrington rotation period of 27.2753 days to define the
angular rotation rate ($\Omega = 2.6662 \times 10^{-6}$ rad s$^{-1}$),
we produced a total of 8727 field lines separated in azimuthal angle
by $\Delta \phi = 0.00917\arcdeg$.

\begin{figure}
\epsscale{1.10}
\plotone{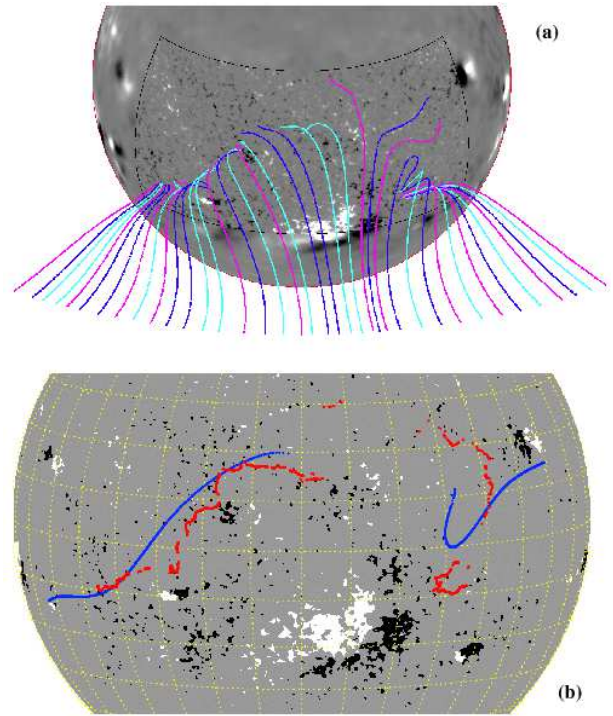}
\caption{(a) View from above the ecliptic plane of PFSS field
lines traced from the source surface down to the photosphere.
Central regions of the high-resolution SOLIS magnetogram are
embedded in a lower-resolution synoptic magnetogram for CR 2007.
(b) Saturated grayscale version of the high-resolution magnetogram
with lines of heliographic longitude and latitude shown at
{10\arcdeg} separation (yellow dotted curves).
Also shown are footpoint locations of the 8727 high-resolution
field lines (red points) and of field lines traced from the same
starting locus, but using low-resolution ($\ell_{\rm max}=9$)
WSO synoptic data (blue points).
\label{fig01}}
\end{figure}

Figure \ref{fig01} shows two illustrative views of the field lines
and footpoints.
A subset of the field lines is shown from a viewpoint above the
ecliptic plane in Figure \ref{fig01}(a), along with the
high-resolution SOLIS magnetogram embedded in the low-resolution
synoptic magnetogram.
Figure \ref{fig01}(b) shows the full set of 8727 footpoints on a
saturated version of the high-resolution magnetogram (i.e.,
three gray levels only) from the viewpoint of an observer on Earth.
We also plot an alternate locus of footpoints that was computed from
a low-resolution PFSS model made with synoptic magnetogram data from
the Wilcox Solar Observatory \citep[WSO; see][]{HS86}.
The published PFSS coefficients for CR 2007 were constructed with
maximum order $\ell = 9$ in the spherical harmonic expansion.
Figure \ref{fig01}(b) shows that at this time the heliographic
latitude $B_{0}$ was near its annual maximum of {+7.2\arcdeg}.

The magnetic polarities in the QS footpoint area of interest
were largely mixed and balanced, but there was a slight preponderance
of negative polarity in this region of the photosphere.
An approximate measurement of the flux imbalance fraction $\xi$
(i.e., the ratio of net flux density to absolute unsigned flux
density) yielded $\xi \approx 0.15$, which is well
within the range of values expected for QS regions
\citep{WS04,Hg08}.
The slight imbalance toward negative polarities was amplified at
larger distances, such that at the PFSS radius the entire
{80\arcdeg}~wide sector shown in Figure \ref{fig01}(a) was
uniformly negative in its polarity.
Thus, each of the 8727 open field lines exhibited a negative polarity
at their footpoints as well.

If the full set of field lines was mapped down {\em radially}
from their initial locations to the solar surface, the horizontal
separation of each neighboring pair of footpoints would be
approximately 110 km.
It is clear from Figure \ref{fig01} that the actual distribution
of all 8727 footpoint separations is quite broad.
Clumps of closely packed footpoints are separated by wider
jumps, the latter occurring in regions where the large-scale
polarity changes.
The distribution of neighboring footpoint separations spans
several orders of magnitude, with minimum, mean, and maximum
values of 5.2, 281, and 249,000 km, respectively.
The distribution has a median value of 40 km, and it is skewed
such that 75\% of the neighbor separations are less than 100 km.

In contrast to the statistics of neighboring footpoint
separations given above, we note that the SOLIS pixel size is
roughly 820 km on the solar surface.
Thus, our reconstructed magnetic field lines often tend to
oversample the information available in the photospheric data.
However, there is evidence that the overlying coronal magnetic
field may exhibit topological features---such as nulls, separatrices,
and quasi-separatrix layers (QSLs)---that exist on spatial scales
smaller than those of the driving sources
\citep[e.g.,][]{Cl05b,An07,An11}.
It is not clear how much extra resolution is needed in order to
resolve these features properly and to produce models that contain
the full diversity of flux-tube expansion properties in the
open coronal field, so we chose to over-resolve
\citep[see also][]{Gf12}.

\begin{figure}
\epsscale{1.00}
\plotone{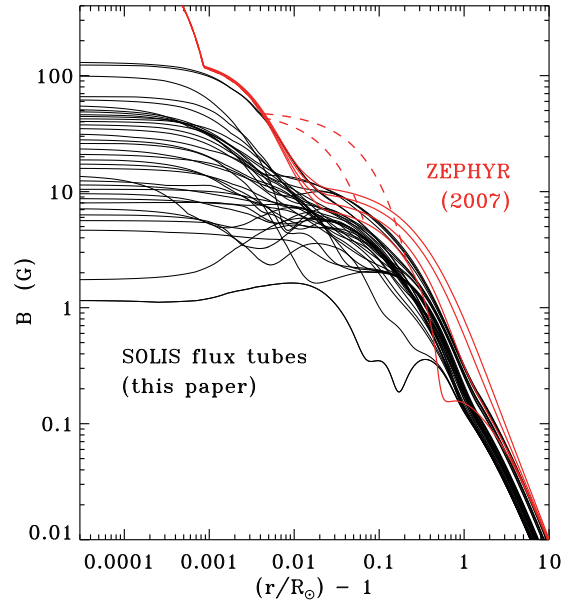}
\caption{Radial dependence of magnetic field strengths along
a subset of field lines traced in the SOLIS PFSS model (black
solid curves).
We compare these with the magnetic field distributions used by
\citet{CvB07} for ZEPHYR solar wind models.
Solid red curves show a range of latitudes in the quiet solar
minimum configuration of \citet{Bana}.
Dashed red curves show open flux tubes assumed to be rooted
in strong active regions.
\label{fig02}}
\end{figure}

Figure \ref{fig02} shows the radial dependence of magnetic field
strength along a subset of the high-resolution PFSS field lines.
Above the source surface radius of 2.46 $R_{\odot}$, the radial
field strength is extrapolated as $B_{r} \propto r^{-2}$.
The models described in this section contain no explicit azimuthal
field component $B_{\phi}$ above the source surface.
At 1~AU, the modeled field strengths range between about $10^{-5}$
and $2.5 \times 10^{-5}$ G, which is about a factor of two
smaller than is generally measured in the ecliptic \citep{MN90}.
However, the two-dimensional models described in Section \ref{sec:cir}
contain both a self-consistent Parker spiral field (i.e.,
$B_{\phi} \neq 0$) and intermittent enhancements of $B_r$ due to
stream interactions.

For the full set of 8727 modeled field lines, the distribution of
field strengths at the photospheric lower boundary has a minimum
and maximum of 1.16 and 131 G, respectively.
The median photospheric field strength is 24.1 G, the mean is
31.3 G, and the standard deviation about the mean is 26.3 G.
Figure \ref{fig02} shows 36 field lines from the PFSS model:
34 chosen at random, plus the ones with minimum and maximum
field strengths at the photosphere.
In Figure \ref{fig02} we also plotted the radial dependence of
field strengths for the flux tubes used in the solar wind
models of \citet{CvB07}.

We use the modeled radial distributions of field strength to
extract a convenient scalar measure of each flux tube's
cross-sectional expansion in the corona.
In order to quantify the degree of superradial expansion, we define
the Wang-Sheeley-Arge (WSA) expansion factor $f_{\rm ss}$ as
\begin{equation}
  f_{\rm ss} \, = \,
  \frac{(B r^{2})_{\rm base}}{(B r^{2})_{\rm ss}} \,\, ,
  \label{eq:fss}
\end{equation}
where the subscript ``base'' refers to the photospheric lower
boundary of the PFSS extrapolation and ``ss'' refers to the
source surface radius of 2.46 $R_{\odot}$.
Much like the footpoint locations and photospheric field strengths,
we find that $f_{\rm ss}$ varies intermittently with longitude.

\begin{figure}
\epsscale{1.00}
\plotone{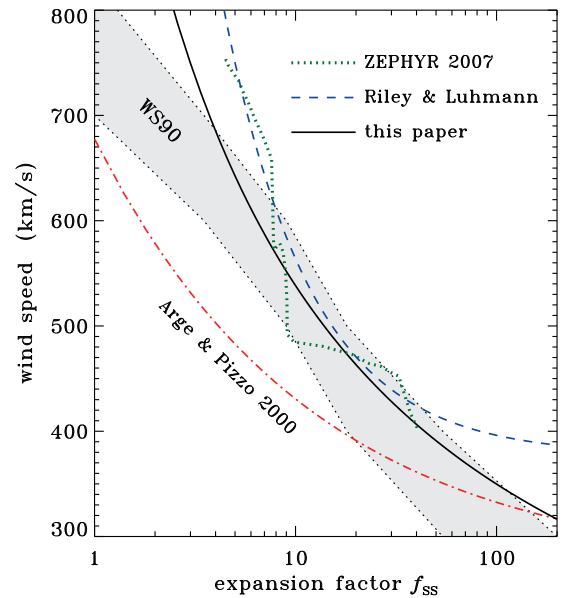}
\caption{Empirical relationships between the flux-tube expansion
factor $f_{\rm ss}$ and the terminal solar wind speed $u_{\infty}$.
Published parameterizations of \citet{AP00} (red dot-dashed
curve) and \citet{RL12} (blue dashed curve) are compared with
the original data-ranges given by \citet{WS90} (gray region),
solar wind model results of \citet{CvB07} (green dotted curve),
and Equation (\ref{eq:ourwsa}) (black solid curve).
\label{fig03}}
\end{figure}

There are several independent calibrations of the well-known
empirical relationship between $f_{\rm ss}$ and the solar wind speed
\citep{WS90,AP00}.
These parameterizations depend on the spatial resolution and
noise level of the photospheric magnetogram data, as well as on
the method used to extrapolate the field
\citep[see also][]{We97,Ow05,Mg11,RL12}.
Figure \ref{fig03} compares several of these parameterizations
with one another.
Note that the \citet{RL12} curve largely reproduces the range
of values reported by the \citet{We97} revaluation of the
original \citet{WS90} analysis.
Figure \ref{fig03} also shows a ``concordance'' WSA relationship
that we constructed to fall within the range of the existing
correlations.
This relationship is given by
\begin{equation}
  u_{\infty} \, = \,
  \frac{2300}{\ln f_{\rm ss} + 1.97}
  \,\,\,\,\, \mbox{km} \,\,\, \mbox{s}^{-1} \,\, ,
  \label{eq:ourwsa}
\end{equation}
where $u_{\infty}$ is the terminal or asymptotic solar wind speed.
We do not take into account the other commonly used variable
parameter of the WSA model: the transverse angular distance
$\theta_{b}$ between the field line of interest and the
boundary of the nearest large-scale coronal hole.
When using this parameter, it is often found that the empirical
wind speed becomes independent of this angle for
$\theta_{b} \gtrsim {7\arcdeg}$ \citep[e.g.,][]{Mg11}.
The patch of QS in which we model the footpoints of slow solar
wind appears to be far enough away from any large coronal holes
that these field lines are probably insensitive to $\theta_{b}$.

\begin{figure}
\epsscale{1.11}
\plotone{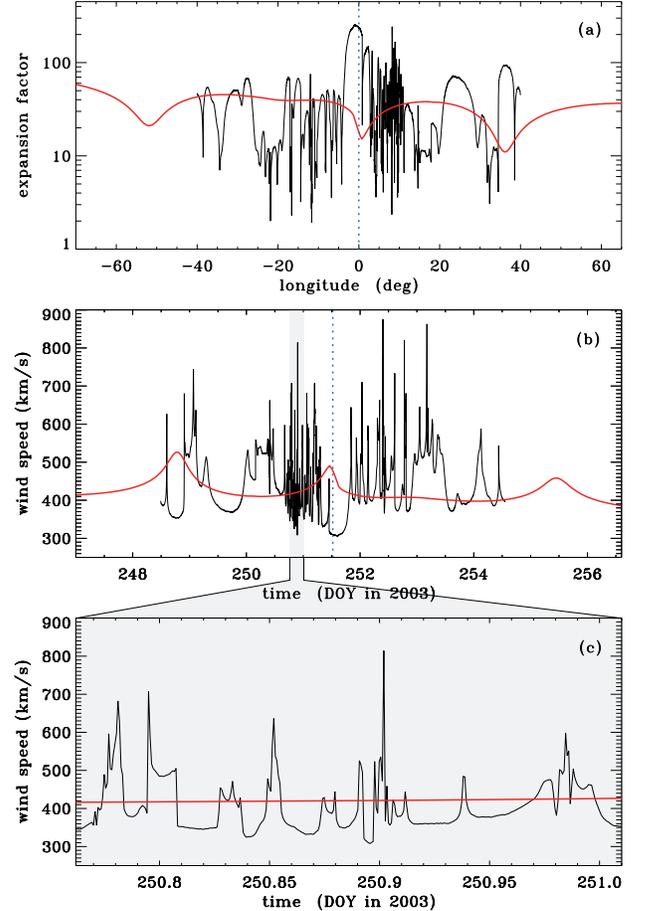}
\caption{(a) Longitudinal dependence of $f_{\rm ss}$ computed
from the high-resolution SOLIS model (black solid curve) and the
low-resolution WSO model (red solid curve).
(b) Empirically estimated wind speed $u_{\infty}$ shown as a
function of time in day-of-year (DOY) units, with same curve
types as in (a).
(c) Six-hour expanded subset of the data shown in (b).
The central meridian location of the SOLIS high-resolution
magnetogram is shown in (a) and (b) with a blue dotted line.
\label{fig04}}
\end{figure}

Figure \ref{fig04} shows the longitudinal dependence of the
flux-tube expansion factor and empirical wind speed $u_{\infty}$
for all 8727 modeled field lines.
We also show corresponding parameters of the PFSS model computed
from the low-resolution WSO magnetograms.
Figure \ref{fig04}(a) plots $f_{\rm ss}$ as a function of a
longitudinal angle ($\phi - \phi_{0}$) normalized by the
central-meridian Carrington longitude from the high-resolution
SOLIS measurement of 2003 September 4.
Figure \ref{fig04}(b) shows $u_{\infty}$, computed from
Equation (\ref{eq:ourwsa}), as a function of time $t$ estimated
for solar wind flux tubes to rotate past the Earth.
For this plot, the conversion between longitude and time is
given by
\begin{equation}
  \phi - \phi_{0} \, = \, -\Omega \left[ t -
  \left( t_{0} + \frac{\Delta r}{\langle u \rangle} \right)
  \right] \,\, ,
  \label{eq:phitime}
\end{equation}
where $t_{0}$ is the time that longitude $\phi_0$ was on the
central meridian, and we use
$\Delta r = 214 \, R_{\odot}$ and
$\langle u \rangle = 450$ km s$^{-1}$ to estimate the mean
travel-time from the Sun to 1~AU.
Time is plotted in Figure \ref{fig04} in units of the day of year
(DOY) in 2003, and the modeled {80\arcdeg} of longitude converts
into approximately 6 days of rotation time.
In Figure \ref{fig04}(c) we show a subset of 6 hours to illustrate
that extremely sharp structures in $f_{\rm ss}$ exist in the
PFSS model that may generate similarly sharp structures
in the solar wind (at least close to the Sun).
In Section \ref{sec:cir} we model the transit-time evolution of
the solar wind more accurately.

We emphasize that our high-resolution PFSS extrapolation of the
Sun's magnetic field represents merely a single ``snapshot'' in time.
In reality, the small-scale magnetic carpet \citep{TS98} evolves
over a wide range of timescales and causes the open-field footpoint
locations to move around.
Specifically, observations have indicated that the
``recycling times'' for magnetic flux in the photosphere
and corona probably are between 1 and 10 hours
\citep[e.g.,][]{Cl05a,Hg08}.
Monte Carlo models of the magnetic carpet's evolution also found
a similar range for the mean timescale of the opening up of
closed flux tubes via reconnection \citep{CvB10}.
These times are comparable with how long it takes
the radial projection of a supergranular cell to rotate past a
distant observer (roughly 3--4 hr).
Thus, we do not mean to present the high-resolution PFSS
reconstruction as a true dynamical model, but only as a
representative state of the QS field that the solar wind will
``see'' as it accelerates up through the open flux tubes.

\section{Non-Potential Radial Enhancements to the Field}
\label{sec:nonpot}

The previous section described the straightforward PFSS
extrapolation of measured photospheric fields into the corona.
However, that process does not take into account the full range
of magnetic field variations that we believe exist along these
flux tubes.
In this section we describe several adjustments that are made
to the radial dependence of the magnetic field strength $B(r)$
for the modeled set of time-steady field lines.

First, we recognize that the photospheric footpoints of the
large-scale coronal magnetic field appear to be broken up into
thin flux tubes (i.e., observed widths of order 50--200 km)
that collect in the dark lanes between the $\sim$1000 km
diameter granulation cells.
These flux tubes have field strengths of 1--2 kG and are often
called ``G-band bright points'' (GBPs) because they show up as
bright features in the 4290--4320 {\AA} molecular bands
\citep[e.g.,][]{Bg95,St01}.
Horizontal motions of these features have been used to put
empirical constraints on the photospheric flux of Alfv\'{e}n wave
energy that propagates up into the corona \citep{Nis03,CvB05,Chi12}.
GBPs do not show up individually in the SOLIS magnetograms that
we used to reconstruct the coronal field, so we modify $B(r)$ as
described below to account for their supposed presence.

\begin{figure}
\epsscale{1.00}
\plotone{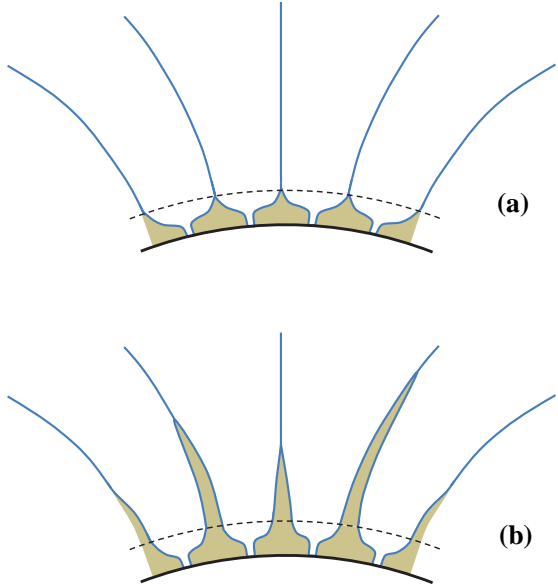}
\caption{Schematic illustration of flux-tube expansion in
the presence of non-potential fields.
In both panels, open flux tube boundaries (blue solid curves)
surround empirically determined field-free regions (gold areas).
(a) For $B_{1} \neq 0$ and $B_{2} = 0$, the only enhancements to
the open field occur below the mid-chromospheric ``merging height''
(dashed curve) of \citet{CvB05}.
(b) Postulated field enhancements in both the photosphere and the
low corona are shown (i.e., $B_{1} \neq 0$ and $B_{2} \neq 0$).
\label{fig05}}
\end{figure}

Figure \ref{fig05}(a) illustrates how the assumption of photospheric
field fragmentation results in narrower flux tubes and higher
field strengths than would be otherwise obtained from the
magnetogram data.
Somewhat crudely, we assume the lower solar atmosphere is divided
into strong-field flux tubes and ``field free'' regions that are
in total pressure equilibrium with one another.
Enhancing the field strength inside the flux tube is equivalent
to assuming a larger field-free volume between the tubes.
Thus, along each flux tube we modify the original PFSS field
$B_{\rm 0}(r)$ by adding two additional components in
quadrature; i.e.,
\begin{equation}
  B^{2} \, = \, B_{0}^{2} + B_{1}^{2} + B_{2}^{2}
  \,\,\, .
\end{equation}
The photospheric GBP enhancement $B_1$ is given by
\begin{equation}
  B_{1}(z) \, = \, B_{\odot} \, \exp \left( - \frac{z}{2 H_{1}}
  \right)  \,\, ,
  \label{eq:B1}
\end{equation}
where $z = r - R_{\odot}$ is the height above the (optical depth
unity) photosphere, and we adopt $B_{\odot} = 1470$ G as a
universal GBP field strength.
The upper photospheric scale height $H_{1} = 120$ km corresponds
to a hydrostatic temperature of approximately 4000 K.
These constants result in a field strength that matches the
photospheric and lower chromospheric parts of the
\citet{CvB05} flux tube model.

There is evidence that additional modifications to a potential
field (i.e., the $B_2$ term above) are needed in the corona.
For example, \citet{CvB05} took account of the large-scale bundling
of fields in the supergranular network of a coronal hole by
constructing a two-dimensional magnetostatic model of the
canopy-like expansion of open flux tubes
\citep[see also][]{Gb76,Gi80,Hk00,AR06}.
The presence of weak-field regions in the internetwork cell centers
(with presumably higher gas pressure than the strong-field network
lanes) modifies the potential field expansion at chromospheric heights.
Similarly, \citet{SvB05} studied the modifications to a potential
field due to the presence of finite gas pressure gradients throughout
the low corona.
They modeled the spatial dependence of the plasma $\beta$ parameter
(i.e., the ratio of gas pressure to magnetic pressure) in a QS region
and found that $\beta \approx 1$ over much of the volume.
This implies the gas influences the field topology in ways unanticipated
by the potential field model, which implicitly assumes $\beta \ll 1$.

It is well known that active regions frequently contain large-scale
currents that give rise to twisted, sigmoidal field lines
\citep[e.g.,][]{Ga87,Cn99,Sj05}.
These structures are believed to be the result of a combination
of surface shear motions and the emergence of new flux from below
the solar surface.
However, because these effects also occur elsewhere on the Sun,
it is likely that other regions (including the QS footpoints of open
flux tubes) exhibit currents and non-potential fields on a wide
range of spatial scales \citep[see][]{Ab07,Zh09,Ye10,Rd11,MvB12}.
These effects give rise to increased ``fibril'' type complexity to
the field.
Their presence is likely to increase the field strength in the low
corona and thus shrink the volume of any given open flux tube that
traverses the non-potential region
(see Figure \ref{fig05}(b)).
Other physical effects that may contribute to modifying the
radial dependence of $B$ include
pervasive chromospheric upflows \citep{MD09},
rotational supergranule motions \citep{ZL11},
and loop footpoint asymmetries that give rise to rapid
unresolved motions \citep{Wg10}.

In the absence of a clear-cut method of modeling non-potential
field enhancements in the low corona, we adopt a similar
hydrostatic radial dependence as was used in Equation (\ref{eq:B1}).
The added magnetic field component is given by
\begin{equation}
  B_{2}(z) \, = \, B_{0}(0) \, \exp \left( - \frac{z}{2 H_{2}} \right)
  \,\, ,
  \label{eq:B2}
\end{equation}
where an approximate scale height
$H_{2} = 2 k_{\rm B} T_{2} / (m_{p} g)$ is defined by an arbitrary
effective temperature $T_2$, the Boltzmann constant $k_{\rm B}$,
the proton mass $m_p$, and the Sun's surface gravitational
acceleration $g$.\footnote{%
We simplified the definition of $H_2$ by assuming a fully ionized
hydrogen plasma; i.e., we did not include the impact of helium and
other heavy ions on the mean molecular weight, and we did not
include partial ionization effects.}
Because the primary goal of this non-potential enhancement is to
``stretch out'' the pre-existing field, it is normalized using the
PFSS lower boundary condition $B_{0}(0)$ of that particular
field line.
The effective temperature $T_2$ is a purely empirical parameter
that characterizes the radial extent of the $B_2$ field modification.
The only way that $T_2$ would be related to an actual coronal
temperature would be if gas pressure effects were the primary
cause of the non-potential enhancement \citep[see, e.g.,][]{SvB05}.

\begin{figure}
\epsscale{1.00}
\plotone{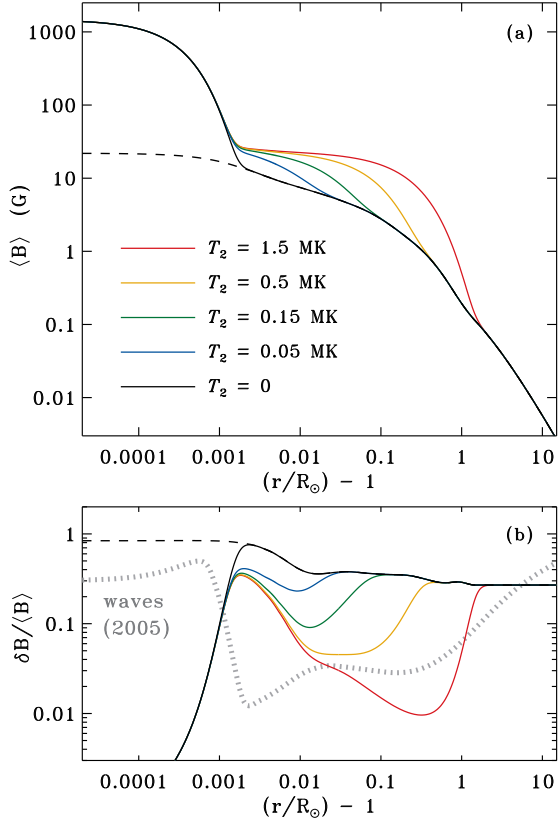}
\caption{Radial dependence of (a) mean magnetic field strengths and
(b) ratios of rms field strength fluctuations to the mean field
strengths.
The unmodified $B_0$ model (black dashed curve) is shown with a
model that includes only the $B_1$ enhancement (black solid curve)
and others that also contain the $B_2$ coronal enhancement
(color solid curves, see labels for $T_2$ values).
Also shown in (b) is the ratio of rms to mean field strength in
the non-WKB Alfv\'{e}n wave model of \citet{CvB05} (gray dotted
curve).
\label{fig06}}
\end{figure}

Figure \ref{fig06}(a) illustrates the magnetic field enhancements
described above using a mean PFSS model.
The unmodified field strength $B_{0}$, shown by the dashed curve,
was produced by forming the average of $\ln B_{0}$ at each height
and taking the exponential of the result.
The photospheric enhancement $B_1$ dominates the modified field
strength at heights $z \lesssim 0.002 \, R_{\odot}$.
The other curves show the result of varying the effective
scale-height temperature $T_2$ in Equation (\ref{eq:B2}).
Values of $T_2$ between about 0.05 and 0.5 MK represent mild
enhancements to the field strength in the low corona that are
consistent with the flux-tube constrictions illustrated in
Figure \ref{fig05}(b).
The red curve corresponding to the largest field strength in
Figure \ref{fig06}, which was computed for $T_{2} = 1.5$ MK, is
probably beyond the realm of physical realism.
Generally, we do not expect the non-potential effects described
above to extend as far up as $z \gtrsim 0.5 \, R_{\odot}$
in a QS region.
We also do not believe that using a value of $T_2$ that is
of the same order of magnitude as the expected peak coronal
temperature is especially realistic, either.
We show the $T_{2} = 1.5$ MK curve mainly for the sake of
completeness.

In Figure \ref{fig06}(b) we show a measure of the statistical
variability of the field strength in the whole set of 8727 SOLIS
flux tubes.
The plotted quantity is the ratio of the standard deviation in
$B$ to its mean value at a given height.
As above, the dashed line shows the unmodified field strength
$B_0$, which reaches a maximum ratio of $\sim$0.84 in the photosphere
and declines to about 0.27 at the source surface.
When the $B_1$ and $B_2$ enhancements are added there is
necessarily a smaller degree of variability in $B$, especially
for the strong photospheric component at low heights.
In the corona, the ratio gets smaller for two reasons of nearly
comparable magnitude to one another:
(1) the mean field strength increases, and
(2) the standard deviation decreases.
We also compare these ratios of relative variability to a
representative model of Alfv\'{e}n wave variability from \citet{CvB05}.
The plotted quantity is the ratio of the root mean squared (rms)
Alfv\'{e}nic fluctuation amplitude $\delta B$ to the background
field strength $B_0$.
Of course, this is somewhat an ``apples to oranges'' comparison
because the Alfv\'{e}nic variability represents propagating
transverse waves and the other curves describe a collection of
{\em static spatial} fluctuations.
Nonetheless the overall similarity of the two ratios indicates
that the two types of variations may be of comparable importance
in the low corona.

We chose not to include the non-potential field enhancements
$B_1$ and $B_2$ in our calculations of the superradial expansion
factor $f_{\rm ss}$ for each field line.
These enhancements are included mainly for the benefit of the
ZEPHYR simulations, which make use of the Alfv\'{e}n speed profile
in the lower atmosphere to translate photospheric velocity
amplitudes into wave energy fluxes.
If we had inserted the modified field strengths in
Equation (\ref{eq:fss}) to rescale $f_{\rm ss}$, the basal field
strengths would all be nearly identical to one another due to the
$B_1$ term.
In that case, much of the relevant information about field-line
{\em spreading} in the upper chromosphere and low corona (which
is contained in the measured variability of $B_{0}(0)$) would
have been erased from $f_{\rm ss}$.

\section{Turbulence-Driven Solar Wind Models}
\label{sec:zephyr}

We used the computed magnetic fields $B(r)$ as inputs to a series
of one-dimensional physical models of turbulent coronal heating
and solar wind acceleration.
The steady-state models presented below are numerical solutions to
one-fluid conservation equations for mass, momentum, bulk internal
energy, and Alfv\'{e}n wave energy.
\citet{CvB07} described these equations and outlined the
computational methods used to solve them using a computer code
called ZEPHYR.
The models presented below were calculated with a slightly modified
version of the original ZEPHYR code.
In addition to several algorithmic improvements that were needed
to allow the code to read in large numbers of flux tubes at a time,
we made three modifications that affected the results:
\begin{enumerate}
\item
We changed the value of the coefficient that multiplies
\citeauthor{Ho74}'s (\citeyear{Ho74}, \citeyear{Ho76}) prescription
for free-streaming conductive heat flux in the collisionless
heliosphere.
\citet{CvB07} used $\alpha_{c} = 4$, as originally suggested by
Hollweg, but we reduced it to $\alpha_{c} = 1$ based on a recent
analysis of electron heat flux measurements in the fast solar wind
\citep[see][]{Cm09}.
\item
We reduced the photospheric boundary condition on the energy flux
of longitudinal acoustic waves from
$10^8$ to $10^6$ erg s$^{-1}$ cm$^{-2}$.
The lower value gave a more realistic height for the transition
region (TR) between the chromosphere and corona than did the higher
value.
Figure~8 of \citet{CvB07} showed that a larger acoustic wave
pressure in the chromosphere gives rise to a larger density
scale height and thus causes the ``critical'' density for
runaway radiative instability to occur at a larger height.
Our adopted value of the acoustic wave flux, in combination with
the \citet{CvB07} choice for the photospheric Alfv\'{e}n wave
amplitude ($v_{\perp} = 0.255$ km s$^{-1}$), was held fixed
for all of the models discussed below.
\item
We used the modified version of the numerical relaxation method for
the internal energy equation described by \citet{Cr08}.
We also reduced the initial value of the minimum undercorrection
exponent $\epsilon_0$ from 0.17 to 0.10.
These changes gave rise to more robust convergence of the coronal
and heliospheric temperature $T(r)$ to its steady-state solution.
\end{enumerate}
It should also be emphasized that ZEPHYR code makes use of the full
radial dependence of $B(r)$ and does not depend on the spatial
resolution of observations that can affect the normalization of
the expansion factor $f_{\rm ss}$.

Prior to computing models for a large number of flux tubes from
the SOLIS reconstruction, we performed a limited parameter study
to explore the effects of varying the non-potential field
enhancement $B_2$.
We began with the mean-field models shown in Figure \ref{fig06}(a)
and produced a finer grid that varied the $T_2$ parameter from
0 to 1.5 MK in increments of 0.025 MK.
These models all used identical lower boundary conditions in the
photosphere, and only differed in their tabulated $B(r)$ field
strengths.
Of those 61 models, 52 of them converged successfully to a
steady-state solution that satisfied internal energy conservation
to within 5\% accuracy.\footnote{%
In other words, these models exhibited final values of the
convergence parameter $\langle \delta E \rangle$ less than 0.05.
This parameter was defined in Equation (63) of \citet{CvB07}
and its iterative convergence was illustrated in Figure~4 of
that paper.}
The other 9 models corresponded to values of $T_{2} > 0.9$ MK,
which we suspect is outside the realm of physical realism
for the non-potential field enhancements.

\begin{figure}
\epsscale{1.00}
\plotone{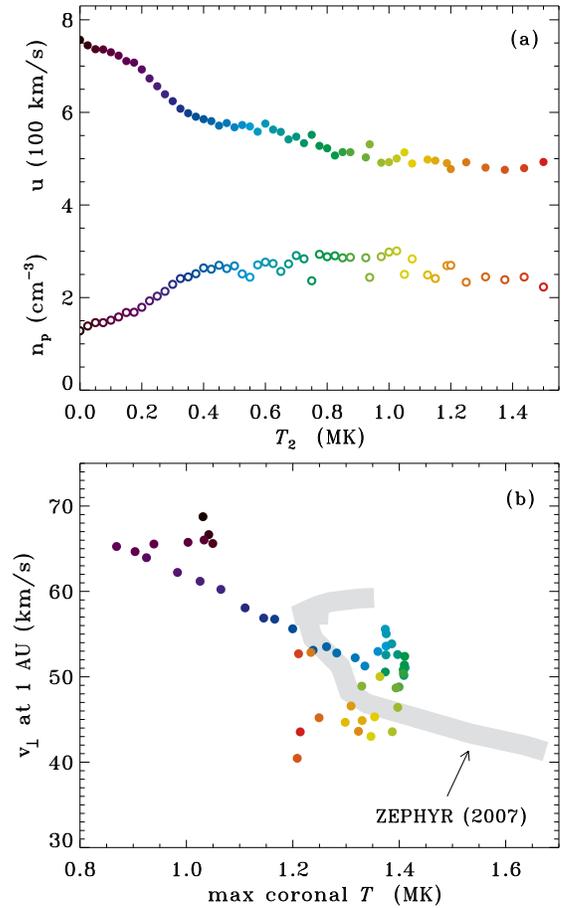}
\caption{Results of mean-field ZEPHYR solar wind models for a
range of $T_2$ parameters.
(a) Wind speed at 1~AU in units of 100 km s$^{-1}$ (solid symbols)
and proton number density at 1~AU (open symbols).
Symbol colors are mapped directly to the $T_2$ values shown
on the horizontal axis.
(b) Alfv\'{e}n wave velocity amplitude at 1~AU plotted vs.\  the peak
coronal temperature for the mean-field models (symbols with
same $T_2$ colors as in panel (a)) and for the models presented by
\citet{CvB07} (gray region).
\label{fig07}}
\end{figure}

In Figure \ref{fig07} we show several summary parameters of the
successful subset of mean-field ZEPHYR models that varied the $T_2$
parameter.
As the non-potential field strength in the low corona is increased,
the wind speed $u$ at 1~AU decreases and the proton number density
$n_p$ at 1~AU increases.
Figure \ref{fig07}(a) shows that most of this variation occurs as
$T_2$ is increased from 0 to 0.5 MK.
Subsequent increases from 0.5 to 1.5 MK do not appear to produce
substantial changes in the modeled solar wind.
Figure \ref{fig07}(b) plots the maximum coronal temperature
versus the Alfv\'{e}n wave velocity amplitude $v_{\perp}$ at 1~AU.
One can think of this value of $v_{\perp}$ as a ``residual''
amplitude since it is the end result of wave dissipation that
occurred in the corona and inner heliosphere.
It makes sense that larger coronal temperatures correspond to
smaller values of $v_{\perp}$, since more coronal heating is
consistent with more damping of MHD turbulence.

We can explain the variations shown in Figure \ref{fig07}
by noticing that the turbulent heating rate $Q$ is often close to
being linearly proportional to the background field strength $B$ in
the upper chromosphere and low corona.
\citet{Cr09} showed that this proportionality is exact in cases
where both the thin flux-tube relation ($B \propto \rho^{1/2}$)
and sub-Alfv\'{e}nic wave action conservation
($v_{\perp} \propto \rho^{-1/4}$) are valid.
Thus, when the field strength $B_2$ is increased, the amount of
heat deposited at the coronal base increases as well.
In direct response, the corona's base pressure increases, as does
the solar wind's mass loss rate \citep[see][]{Hm82,Wi88,CS11}.
This explains why the peak coronal temperature and the proton number
density at 1~AU both increase when $T_2$ is increased.
The decrease in $v_{\perp}$ at 1~AU was explained above as a result
of the increased wave damping that goes along with stronger coronal
heating.

The variation in the wind speed $u$ in Figure \ref{fig07} can be
understood as a result of the presence of ponderomotive
wave-pressure acceleration \citep{J77,He83}.
When the relative strength of coronal heating is low (i.e., for
the smallest values of $T_2$), the wind is driven primarily by
wave pressure and not gas pressure.
In that regime, there is a higher outflow speed when there is a
stronger (less damped) population of Alfv\'{e}n waves both in
the corona and at 1~AU.
Figure \ref{fig07}(b) shows that when $T_2$ is smaller than
about 0.25 MK, the corona is heated to peak temperatures below 1~MK
and the residual wave amplitude at 1~AU is higher than was seen
in the \citet{CvB07} ZEPHYR models.
Thus, we may be able to rule out values of $T_2$ below $\sim$0.25 MK
because of their unrealistic solar wind properties.
If we combine this with the discussion above that appeared to
also rule out values of $T_2$ larger than 1--1.5 MK, this leaves
only a limited range of $T_2$ parameters (i.e.,
$0.25 \lesssim T_{2} \lesssim 0.75$ MK) that may be relevant for
modeling our reconstructed QS flux tubes.

With the above results in mind, we proceeded to create solar wind
models for the SOLIS flux tubes discussed in Section \ref{sec:solis}.
Because the full set of 8727 flux tubes often oversamples the
existing complexity of the reconstructed magnetic field, we selected
a subset of 289 field lines to input to ZEPHYR.
These field lines were chosen by sampling the data more frequently
in regions of rapid longitudinal change than in regions where the
field strength was nearly constant.
Every local maximum and minimum in the longitudinal variation of
$f_{\rm ss}$ (see Figure \ref{fig04}(a)) was represented by at
least one of the 289 resampled points.
We ran ZEPHYR for three distinct cases of non-potential field
enhancement:
(1) $T_{2} = 0$ (i.e., using the photospheric $B_1$ enhancement only),
(2) $T_{2} = 0.2$ MK, and
(3) $T_{2} = 0.5$ MK.
We decided against using even larger values of $T_2$ because
Figure \ref{fig07} showed that the resulting solar wind would
probably not have been significantly different from the
$T_{2} = 0.5$ MK case.

\begin{figure}
\epsscale{1.10}
\plotone{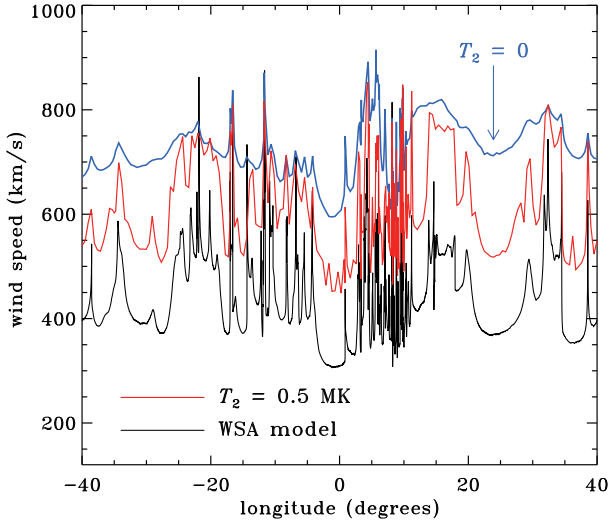}
\caption{Comparison of asymptotic solar wind speeds estimated from
the WSA anticorrelation (lower black curve) with those computed
from ZEPHYR models having $T_{2} = 0$ (upper blue curve) and
$T_{2} = 0.5$ MK (middle red curve).
Longitudes are plotted for the footpoints of the PFSS flux tubes
as in Figure \ref{fig04}(a).
\label{fig08}}
\end{figure}

Figure \ref{fig08} summarizes the results of two of the ZEPHYR models
($T_{2} = 0$, 0.5 MK) by comparing the wind speeds at 1~AU with the
predicted WSA wind speed from Equation (\ref{eq:ourwsa}).
The intermediate case of $T_{2} = 0.2$ MK exhibits plasma properties
that lie between the two plotted models.
Note that an overall WSA-type anticorrelation between $u$ and
$f_{\rm ss}$ appears to be upheld for the numerical models.
As was discussed above, the $T_{2}=0$ models do not contain enough
turbulent heating to give rise to coronal temperature maxima above
1 MK.
Thus, the resulting lack of wave dissipation gives rise to ``too much''
wave pressure acceleration and a high-speed, low-density wind.
The case of $T_{2} = 0.5$ MK gives rise to wind speeds at 1~AU
between 448 and 853 km s$^{-1}$, which is a bit higher than the
observed range of speeds but is much more realistic than the
predictions of the $T_{2}=0$ model.

\begin{figure}
\epsscale{1.10}
\plotone{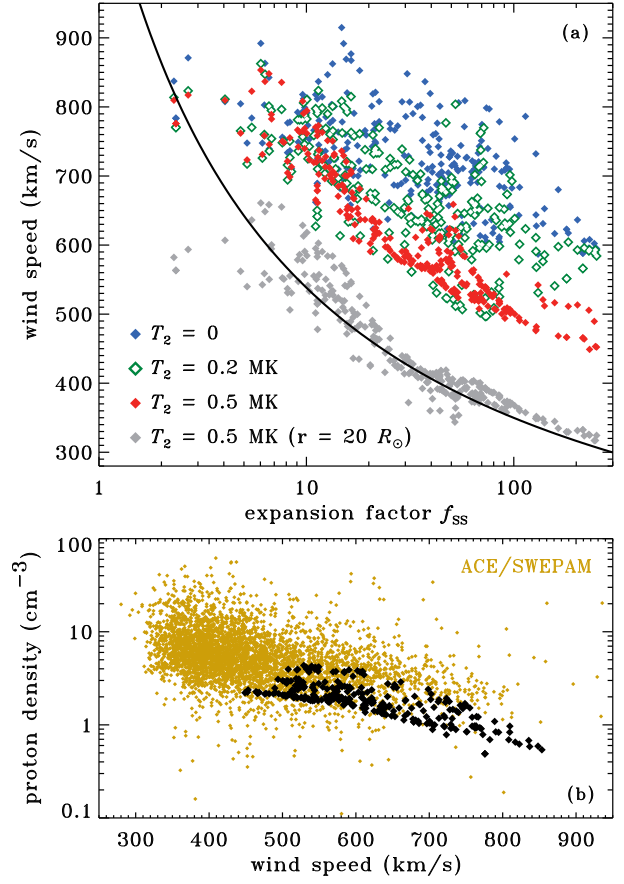}
\caption{(a) Solar wind speeds plotted vs.\  $f_{\rm ss}$
compared with Equation (\ref{eq:ourwsa}) (black solid curve).
Speeds at 1~AU are shown for ZEPHYR models having
$T_{2} = 0$ (blue filled symbols),
$T_{2} = 0.2$ MK (green open symbols), and
$T_{2} = 0.5$ MK (red filled symbols).
Speeds at $r = 20 \, R_{\odot}$ are also shown for the
$T_{2} = 0.5$ MK model (gray filled symbols).
(b) Proton number densities $n_p$ at 1~AU plotted vs.\  wind
speeds for the $T_{2} = 0.5$~MK model (large black symbols) and
for {\em{ACE}}/SWEPAM data (small gold symbols).
\label{fig09}}
\end{figure}

In Figure \ref{fig09}(a) we examine in more detail how the models
appear to follow WSA-like relationships between wind speed and
expansion factor.
Although there is substantial scatter, each set of ZEPHYR results
does seem to be centered around a relationship reminiscent of
Equation (\ref{eq:ourwsa}).
Models with lower values of $T_2$ correspond to larger normalization
offsets in the wind speed.
For the $T_{2} = 0.5$ MK case, we show the modeled wind speeds
measured at a heliospheric distance of $r = 20 \, R_{\odot}$ in
addition to those at 1~AU.
At the former distance the wind speed in each flux tube is roughly
75\% of its asymptotic terminal speed in interplanetary space.
Those speeds agree remarkably well with the curve corresponding to
Equation (\ref{eq:ourwsa}).
This is consistent with the fact that the empirical WSA relationship
is often applied as a lower boundary condition (typically at
distances of order 20 $R_{\odot}$) for global simulations of the
inner heliosphere \citep[e.g.,][]{Le09,Mg11}.

It is difficult to pin down a simple explanation for the emergence
of clear WSA anticorrelations in Figure \ref{fig09}(a).
There have been various attempts to explain how an Alfv\'{e}n wave
driven wind should exhibit this kind of relationship between wind
speed and expansion factor \citep{Kv81,WS91,Sz06}.
These explanations tend to involve differences in the radial
evolution of wave energy flux along different field lines---i.e.,
larger wave fluxes at the critical point correspond to both
greater acceleration and smaller values of $f_{\rm ss}$.
This must be happening to some extent, but the ZEPHYR models also
include non-WKB reflection, self-consistent wave damping, and wind
acceleration from both gas pressure and wave pressure.
These effects are linked to one another via a number of nonlinear
feedbacks, so we do not yet know the precise chain of events
that gives rise to the emergent distribution of wind speeds.
Future work \citep[e.g.,][]{WC12} will aim to determine whether
these interactions can be understood from the standpoint of more
basic scaling relations.

Figure \ref{fig09}(b) shows the anticorrelation between wind speed
and proton number density at 1~AU.
In addition to the $T_{2} = 0.5$ MK model results, we also show
hourly-averaged data measured by the
Solar Wind Electron, Proton, and Alpha Monitor (SWEPAM) instrument
on the {\em Advanced Composition Explorer} ({\em{ACE}}) spacecraft
\citep{Mc98}.
The data displayed are a relatively sparse selection of approximately
4700 data points spread out in time between the years 1998 and 2006.
Even though the models do not extend to the lowest wind speeds and
highest densities seen in the low-latitude heliosphere, the
overall sense of agreement between the models and the measurements
is good.

\begin{figure}
\epsscale{1.00}
\plotone{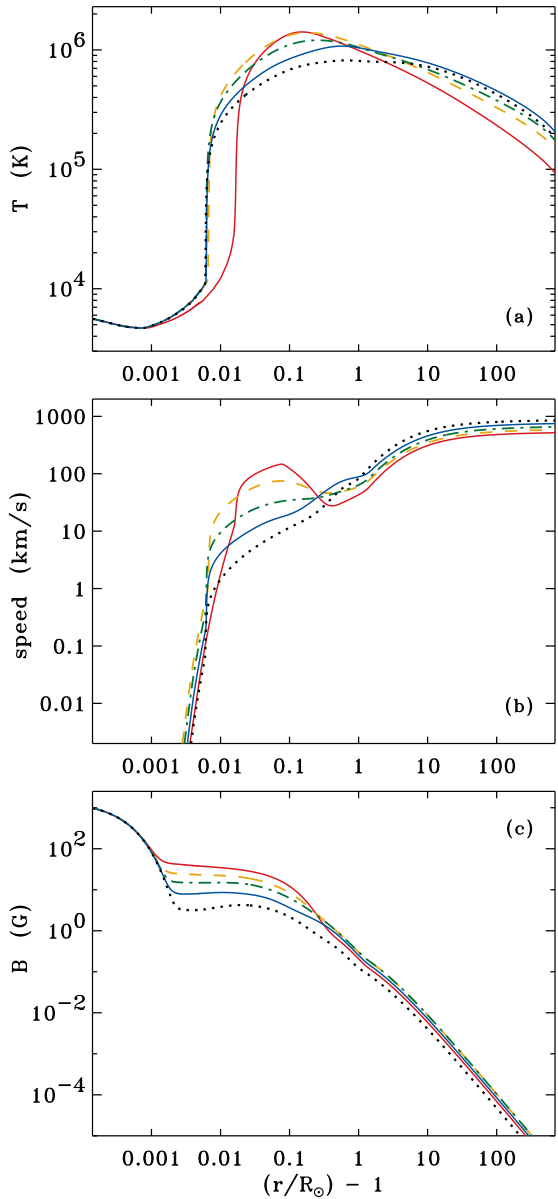}
\caption{Radial dependence of solar wind parameters for a
selection of five flux tubes extracted from the ZEPHYR models
computed with $T_{2} = 0.5$~MK.
Shown are: (a) one-fluid temperatures, (b) radial wind speeds,
and (c) magnetic field strengths vs.\  height above the
solar photosphere.
Line styles and colors serve merely as consistent identifying
labels for the flux tubes.
\label{fig10}}
\end{figure}

To illustrate the plasma properties that appear in the
ZEPHYR models, we show in Figure \ref{fig10} the temperatures,
outflow speeds, and magnetic field strengths for a selection of
five flux tubes from the $T_{2} = 0.5$ MK model.
In the low corona, the magnetic field strengths in these flux tubes
vary over about an an order of magnitude (from 4 to 40 G).
However, Figure \ref{fig10}(c) shows that $B(r)$ does not vary
substantially in the photosphere and chromosphere
($z \lesssim 0.002 \, R_{\odot}$) nor in the extended corona
and heliosphere ($z \gtrsim 0.5 \, R_{\odot}$) of these models.
As described above, stronger coronal fields give rise to more heating
below the critical point, which in turn produces a hotter, denser,
and ultimately slower wind \citep[see also][]{LH80,Pn80,Le82}.
Figure \ref{fig10}(b) shows that the hotter models undergo more
intense ``bursts'' of solar wind acceleration in the low corona,
but their large-scale outflows end up being slower.

Figure \ref{fig10}(a) shows that the model receiving the largest
degree of coronal heating has a higher TR than the other models.
This occurs because that model's intense wave pressure in the
upper chromosphere produces a larger pressure scale height and a
shallower radial decline in the density.
Thus, the density does not drop to its critical value (for runaway
radiative instability) until it reaches a significantly larger height
than in the other models.
A situation similar to the above case occurs for roughly half of
the full set of 289 ZEPHYR models computed with $T_{2} = 0.5$ MK.
Defining the TR height $z_{\rm TR}$ as the location at which
$T$ first exceeds 0.2 MK, we found that the models bifurcate
cleanly into two groups:
162 of them had $z_{\rm TR} \leq 0.008 \, R_{\odot}$
and 127 of them had $z_{\rm TR} \geq 0.012 \, R_{\odot}$.
As expected, the latter set of models with larger values of
$z_{\rm TR}$ tended to have larger peak temperatures
($T_{\rm max} \gtrsim 1.3$ MK) and lower asymptotic wind speeds
($u_{\infty} \lesssim 600$ km s$^{-1}$) than the other models.
This appears to be a kind of ``bistable'' situation, and it is
possible that a truly time-dependent model would exhibit
oscillations between these two states.
We will study this effect in future work.

\section{Interplanetary Evolution of the Solar Wind}
\label{sec:cir}

Each of the individual ZEPHYR models described above was
computed independently of the others.
However, we know that magnetic flux tubes in the corona and
heliosphere interact with their ``neighbors'' in a variety of ways.
The distinct identity of a given flux tube can be blurred by
time-dependent disruptive effects such as Kelvin-Helmholtz
instabilities \citep[e.g.,][]{Ei99,Ph99,Ve11} or diffusive
field-line shredding due to MHD turbulence \citep{Mu91,Mt95}.
Also, the large-scale structure of the heliosphere is often
dominated by collisions between obliquely oriented streams
that result in CIR-like compressions and rarefactions.

Our goal in this section is to begin taking account of the
multidimensional evolution of solar wind flux tubes as initially
computed from the high-resolution magnetic field models of
Sections \ref{sec:solis}--\ref{sec:zephyr}.
We use the longitudinally resolved set of ZEPHYR plasma properties
as an inner boundary condition to a two-dimensional MHD
calculation in the ecliptic plane.
Thus, we simulate the time-steady effects of CIR formation, but we
do not yet include any smaller scale time-dependent effects (e.g.,
waves, instabilities, or turbulence) on the interplanetary evolution
of these flux tubes.

For a frame corotating with constant angular velocity $\Omega$,
\citet{Wh80} and \citet{Hu93} described
the time-steady MHD conservation equations in the ecliptic
plane ($\theta = \pi/2$).
In general, we solve for the six magnetofluid parameters
$\rho$ (mass density), $v_r$ (radial velocity),
$v_{\phi}$ (azimuthal velocity), $P$ (gas pressure),
$B_r$ (radial field strength), and $B_{\phi}$ (azimuthal
field strength) as a function of $r$ and $\phi$.
Mass conservation is given by
\begin{equation}
  \frac{1}{r^2} \frac{\partial}{\partial r} \left(
  r^{2} \rho v_{r} \right) +
  \frac{1}{r} \frac{\partial}{\partial \phi} \left(
  \rho v_{\phi} \right) \, = \, 0 \,\, .
  \label{eq:cirmass}
\end{equation}
The two nontrivial components of momentum conservation are
\begin{displaymath}
  v_{r} \frac{\partial v_r}{\partial r} +
  \frac{v_{\phi}}{r} \frac{\partial v_r}{\partial \phi} +
  \frac{1}{\rho} \frac{\partial P}{\partial r} +
  \frac{GM_{\odot}}{r^2} \,\,\, = 
\end{displaymath}
\begin{equation}
  \frac{B_{\phi}}{4\pi\rho} ( \nabla \times {\bf B} )_{\theta}
  + \frac{v_{\phi}^2}{r} + \Omega (2 v_{\phi} + \Omega r)
  \label{eq:vrcon}
\end{equation}
\begin{equation}
  v_{r} \frac{\partial v_{\phi}}{\partial r} +
  \frac{v_{\phi}}{r} \frac{\partial v_{\phi}}{\partial \phi} +
  \frac{1}{\rho r} \frac{\partial P}{\partial \phi} =
  -\frac{B_r}{4\pi\rho} ( \nabla \times {\bf B} )_{\theta}
  - \frac{v_{r} v_{\phi}}{r} - 2 \Omega v_{r}
  \label{eq:vpcon}
\end{equation}
where $G$ is the Newtonian gravitation constant, $M_{\odot}$
is the solar mass, and the latitudinal component of the curl
of ${\bf B}$ is given by
\begin{equation}
  ( \nabla \times {\bf B} )_{\theta} \, = \,
  \frac{1}{r} \frac{\partial B_r}{\partial \phi} -
  \frac{\partial B_{\phi}}{\partial r} - \frac{B_{\phi}}{r}
  \,\, .
\end{equation}
The internal energy equation for pressure is
\begin{equation}
  v_{r} \frac{\partial P}{\partial r} +
  \frac{v_{\phi}}{r} \frac{\partial P}{\partial \phi} +
  \rho a^{2} \left[ \frac{1}{r^2} \frac{\partial}{\partial r}
  \left( r^{2} v_{r} \right) + \frac{1}{r}
  \frac{\partial v_{\phi}}{\partial \phi} \right] \, = \, 0 \,\, ,
\end{equation}
and the equation of state $\gamma P = \rho a^{2}$ defines the
single-fluid sound speed $a$.
We choose a constant value for the adiabatic exponent $\gamma$.
The divergence-free constraint on the magnetic field gives rise
to an effective conservation equation for $B_r$, which is
\begin{equation}
  \frac{1}{r^2} \frac{\partial}{\partial r} \left(
  r^{2} B_{r} \right) +
  \frac{1}{r} \frac{\partial B_{\phi}}{\partial \phi} 
  \, = \, 0 \,\, .
  \label{eq:cirmag}
\end{equation}
Lastly, the ``frozen-in'' MHD assumption that the magnetic
field vector remains parallel to the corotating velocity vector
allows us to write $B_{\phi} = v_{\phi} B_{r} / v_{r}$.
Making use of the defining equations for $a$ and $B_{\phi}$
leaves 5 equations for 5 unknowns
($\rho$, $v_r$, $v_{\phi}$, $P$, $B_r$).

The MHD equations as described above involve a number of key
approximations.
Our neglect of explicit time dependence results in the
elimination of MHD waves from the modeled system.
However, these equations do contain the terms needed to model
information propagation on MHD characteristics, which was used
by \citet{Hu93} to successfully predict the formation of shocks
in the outer heliosphere.
Note that we also neglect the source terms in the momentum
and internal energy equations that were required to produce
turbulent coronal heating and wave-pressure acceleration in
the ZEPHYR models.
Thus, we are essentially replacing the effects of turbulent
heating at heights above our inner boundary radius (i.e.,
$r > 20 \, R_{\odot}$) with a simple gas pressure
gradient determined by the adopted value of the polytropic
exponent $\gamma$.

To solve Equations (\ref{eq:cirmass})--(\ref{eq:cirmag}), we
first constructed the $5 \times 5$ matrices that contain the
coefficients multiplying the $r$ and $\phi$ partial derivatives.
Following \citet{Hu93}, we inverted the first of those matrices
and solved for each of the $r$ derivatives by themselves.
To step upward from a lower boundary radius, we implemented the
first-order upwind differencing technique described by
\citet{Pr92} and \citet{RL11}.
The approach of \citet{RL11} was to solve an inviscid Burgers'
equation that corresponds to Equation (\ref{eq:vrcon}) in the
limit of $a \rightarrow 0$ and $V_{\rm A} \rightarrow 0$.
We extended their upwind differencing technique to the full set
of corotating MHD equations, and we used the standard
Courant-Friedrichs-Lewy (CFL) criterion (treating $r$ as the
time-like variable and $\phi$ as the space-like variable) to
set the radial step size.

\begin{figure}
\epsscale{1.00}
\plotone{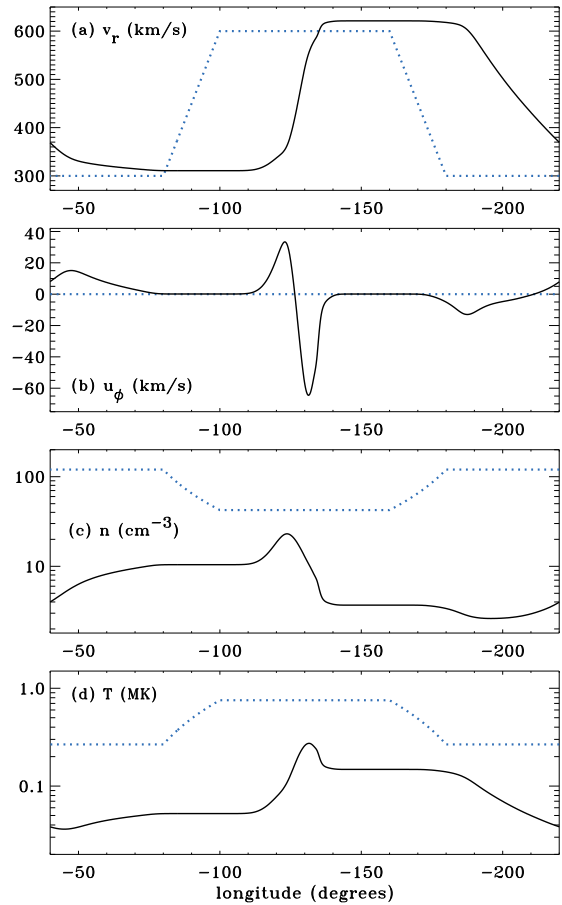}
\caption{Azimuthal profiles of plasma properties computed for
the \citet{Hu93} test problem and shown at $r = 0.3$ AU
(blue dotted curves) and $r = 1$ AU (black solid curves).
Shown are: (a) radial wind speed $v_r$, (b) azimuthal speed
in the inertial frame $u_{\phi}$, (c) proton number density
$n = \rho / m_{p}$, and (c) one-fluid temperature $T$.
The azimuthal angle $\phi$ has been shifted by a constant offset
to have the scale line up with that of Figure 8 of \citet{Hu93}.
\label{fig11}}
\end{figure}

To test our numerical implementation of the MHD equations, we
solved the example problem of a single ``trapezoidal'' shaped
high-speed stream that was posed by \citet{Hu93}.
Starting at a lower boundary radius of 0.3~AU, the wind speed was
doubled (from 300 to 600 km s$^{-1}$) in a {60\arcdeg} wide band
of longitude, the density was correspondingly reduced such
that $\rho \propto v_{r}^{-1.5}$, and the gas pressure was
assumed to remain constant.
This model used an ideal adiabatic value of $\gamma = 5/3$.
Figure \ref{fig11} shows the results of evolving these conditions
from 0.3 to 1 AU, and it compares favorably to the properties
shown in Figure 8 of \citet{Hu93}.
Instead of plotting the azimuthal velocity $v_{\phi}$ in the
corotating frame, Figure \ref{fig11}(b) shows the azimuthal
velocity in the inertial frame,
\begin{equation}
  u_{\phi} \, = \, v_{\phi} + \Omega r  \,\, ,
\end{equation}
which exhibits departures from zero only because of the
stream-stream interactions that grow in magnitude with increasing
heliocentric distance.

Next, we made use of the ZEPHYR solar wind models as inner boundary
conditions for a two-dimensional model of the heliosphere in the
ecliptic plane.
We focused only on the standard parameter choice of $T_{2} = 0.5$ MK
that was used for the models shown in Figure \ref{fig10}.
We extracted the density, wind speed, gas pressure, and magnetic
field strength as a function of $\phi$ at an inner boundary
radius chosen initially to be $r = 20 \, R_{\odot}$.
We chose to set $v_{\phi} = -\Omega r$ at this radius, which is
equivalent to assuming no rotation in the inertial frame
($u_{\phi} = 0$) as is appropriate for a height well above the
solar wind's Alfv\'{e}n point \citep[see, e.g.,][]{WD67}.
Although we assume time-steady corotation of MHD ``patterns'' in
this model system, that is not equivalent to the rigid rotation
of the fluid itself (i.e., assuming $v_{\phi} = 0$ would be
unrealistic).

The fully resolved SOLIS flux tubes subtend only {80\arcdeg} out
of a full {360\arcdeg}, so we interpolated the plasma parameters
linearly throughout the unused {280\arcdeg} of longitude.
We used a fine azimuthal grid with 39279 zones and constant
separation $\Delta \phi = 0.00917\arcdeg$.
We set $\gamma = 1.2$ to account for ongoing gradual heating
in the range of distances to be modeled in the ecliptic plane.
This exponent is roughly consistent with an in~situ temperature
dependence of roughly $T \propto r^{-1/2}$, which is close to
what is seen for the average of proton and electron temperatures
in the inner heliosphere \citep[e.g.,][]{Cm09}.

When running the two-dimensional MHD code for the high-resolution
set of SOLIS flux tubes, we found that in some cases the standard
CFL criterion did not govern the numerical stability of the
upwind differencing scheme.
In these cases, even the use of a radial step size several orders
of magnitude smaller than dictated by CFL would not stabilize the
system's evolution.
We found that Equation (\ref{eq:vpcon}), the azimuthal momentum
conservation equation, was responsible for this instability.
Although \citet{Hu93} found that standard MHD characteristics are
valid for these equations in the supersonic and super-Alfv\'{e}nic
limits, it is possible that the Coriolis and curvature terms in
Equations (\ref{eq:vrcon})--(\ref{eq:vpcon}) modify the system's
effective ``wave speeds'' such that the smaller azimuthal flows
require special treatment.
In practice, we found two alternate ways of producing stable models:
\begin{enumerate}
\item
One solution was to assume that deviations from strict
``Parker spiral'' corotation (i.e., $v_{\phi} = -\Omega r$)
can be neglected.
Figure \ref{fig11}(b) showed that departures from $u_{\phi}=0$
at large-scale stream interfaces are typically subsonic in
magnitude and thus probably unimportant to the overall dynamics
of CIRs.
By making the assumption that $v_{\phi}$ has a known analytic
value, we were able to exclude Equation (\ref{eq:vpcon}) from
the system of equations solved by the code.
The strict upwind differencing method suggested by \cite{RL11}
remained stable for CFL-compliant radial step sizes.
We refer to this as our ``low-diffusion'' model.
\item
Another solution was to solve the $v_{\phi}$ equation using
a method that was both more numerically stable and more diffusive.
We applied the Lax method \citep[see, e.g.,][]{Pr92} to the
solution of Equation (\ref{eq:vpcon}), but we kept the
less diffusive upwind differencing method for the other four
MHD equations.
The inclusion of departures from strict corotation (i.e., allowing
the system to develop longitudinal variations in $v_{\phi}$) also
gives rise to additional {\em physical} diffusion in $\phi$.
Both types of diffusion contribute to the smearing out of sharp
structures in longitude.
Thus, we refer to this case as our ``high-diffusion'' model.
\end{enumerate}
Below we show results from both models.
We believe the true solutions to Equations
(\ref{eq:cirmass})--(\ref{eq:cirmag}) should lie in between
the results of the low and high diffusion cases.

\begin{figure}
\epsscale{1.00}
\plotone{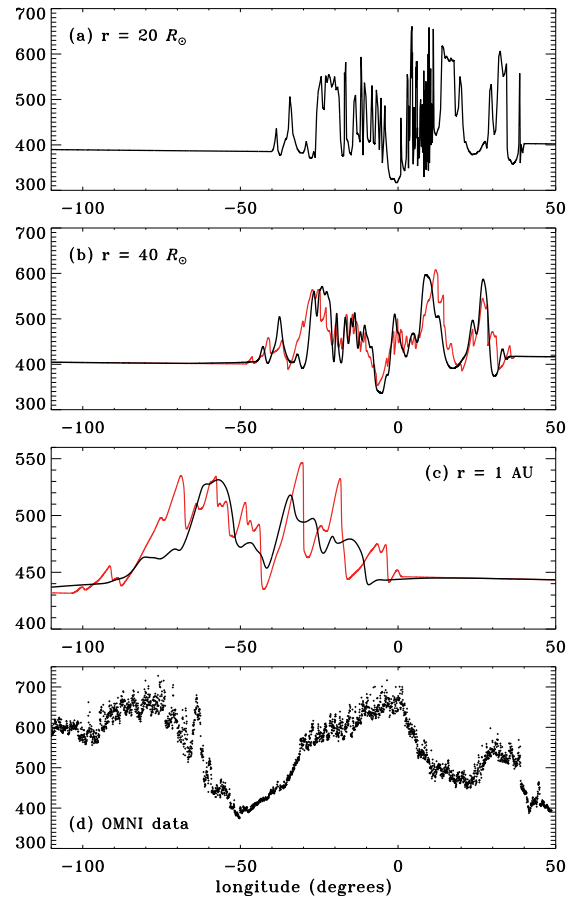}
\caption{Radial wind speeds for MHD models computed from ZEPHYR
solar wind properties.
(a) The inner boundary radius $r = 20 \, R_{\odot}$ contains
the same data for both models.
At distances of (b) $r = 40 \, R_{\odot}$ and
(c) $r = 215 \, R_{\odot} = 1$~AU, the high-diffusion model
(black curves) and low-diffusion model (red curves) differ from
one another.
(d) In~situ data for the modeled time period is shown from the
multi-source OMNIWeb interface.
(One data point is shown every 5 minutes, which is downsampled
from the original time resolution of 1 minute.)
\label{fig12}}
\end{figure}

Figure \ref{fig12} shows how the longitudinal profile of radial
velocity $v_r$ evolves with increasing radial distance.
The ``leftward'' evolution of structures in $\phi$ is due to the
curvature of the overall Parker spiral.
As predicted by \citet{WS03} and others, the strongest and narrowest
velocity peaks at the inner boundary are rapidly smeared out by
interactions with surrounding slow wind.
At the intermediate distance of 40 $R_{\odot}$ shown in
Figure \ref{fig12}(b), the high-diffusion model looks somewhat
like a smoothed version of the low-diffusion model.
However, by the time the streams reach 1~AU (Figure \ref{fig12}(c)),
there has been enough dynamical evolution to make the two models
appear quite different from one another.

Our use of the polytropic exponent $\gamma = 1.2$ resulted in
some extra acceleration of the wind from 20 to 215 $R_{\odot}$.
The mean wind speed at 1~AU is roughly 10\% higher than the
speeds at the inner boundary.
Note that the set of input ZEPHYR models exhibited a slightly
larger degree of acceleration in the inner heliosphere, with their
original speeds at 1~AU being roughly 25\% higher than their speeds
at 20 $R_{\odot}$.
Some of the relative lack of acceleration in the two-dimensional
MHD model could be due to the accumulative loss of kinetic energy
at multiple colliding CIRs.

Figure \ref{fig12}(d) shows the OMNI solar wind speed in the
ecliptic plane (measured mostly by {\em{ACE}}/SWEPAM during this
time period) mapped to longitude using the Carrington rotation
parameters discussed in Section \ref{sec:solis}.
There is not much detailed agreement with the MHD models, but
the persistent local minimum in wind speed centered on
$\phi \approx -40\arcdeg$ may be a relevant similarity between
the data and models.
We should clarify that the MHD models do not include the full
{360\arcdeg} of longitude, so there may be large-scale CIR
interactions with regions {\em outside} the high-resolution set
of flux tubes that affect the solutions inside.
Also, the agreement between global models and in-ecliptic data
often depends crucially on what was assumed for the Sun's
{\em polar fields} \citep[e.g.,][]{Ji11}.
The magnetograms that are used typically for synoptic
reconstructions of the global field (e.g., MDI on {\em SOHO,}
Kitt Peak, WSO) often undergo detailed deprojections and
high-latitude corrections that have not been attempted for the
SOLIS data used here.

\begin{figure}
\epsscale{1.06}
\plotone{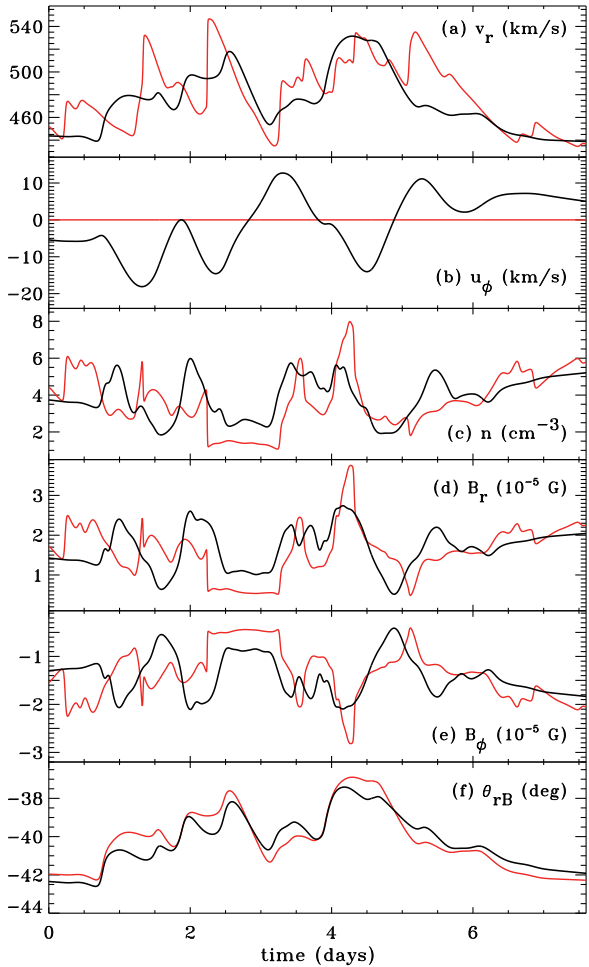}
\caption{Simulated time variability of the low-diffusion (red
curves) and high-diffusion (black curves) models that would be
seen by a spacecraft at 1~AU.
Shown are: (a) radial wind speed $v_r$, (b) azimuthal speed
in the inertial frame $u_{\phi}$, (c) proton number density
$n$, (d) radial field strength $B_r$, (e) azimuthal field strength
$B_{\phi}$, and (f) Parker spiral angle $\theta_{rB}$.
\label{fig13}}
\end{figure}

Figure \ref{fig13} shows a summary of other plasma parameters at
1~AU, here plotted as a function of time as would be measured by
a stationary spacecraft.
We used Equation (\ref{eq:phitime}) to translate longitude into time,
and for simplicity we normalized $t$ to zero at the left edge of
the plot.
Table \ref{table01} gives the mean values of several relevant
parameters at 1~AU for the two models.
One can see more clearly that the low-diffusion model contains
sharper features, often with larger variability, that tend to be
more smeared out in the high-diffusion model.
The tight correlation between $B_r$ and $B_{\phi}$ is a preliminary
indication that the dominant type of magnetic variability is in
the {\em magnitude} of the field and not in its direction.
Also, the modeled fluctuations in gas pressure and magnetic pressure
are largely in phase with one another, and are not {180\arcdeg} out
of phase like one would expect in pressure-balance structures
\citep{Bu90,TM94,VH99}.
These correlations are discussed further in Section \ref{sec:flucts}.
The Parker spiral angle is defined as
$\theta_{rB} = \tan^{-1} (B_{\phi}/B_{r})$.
Figure \ref{fig13}(f) shows that the pattern of longitudinal
variability seen for $\theta_{rB}$ does not depend much on
whether departures from strict corotation are included in
$v_{\phi}$ (as in the high diffusion model) or not (as in
the low diffusion model).

\begin{deluxetable}{ccc}
\tablecaption{Modeled MHD Parameters at 1 AU
\label{table01}}
\tablewidth{0pt}
\tablehead{\colhead{Parameter} &
\colhead{High-diffusion Model} &
\colhead{Low-diffusion Model}}
\startdata
Mean $v_r$ (km s$^{-1}$)    &  474.3   &   481.2 \\
Mean $v_\phi$ (km s$^{-1}$) & --400.2  & --399.0 \\
Mean $n$ (cm$^{-3}$)        &  3.854   &  3.656  \\
Mean $|{\bf B}|$ (nT)       &  2.169   &  2.072  \\
Mean $\beta$ ($P_{\rm gas}/P_{\rm mag}$) &  7.383 & 8.223 \\
\hline
$E_{{\rm K},r} / E_{n}$    & 1.402   & 1.612   \\
$E_{{\rm K},\phi} / E_{n}$ & 0.1407  &   0     \\
$E_{{\rm B},r} / E_{n}$    & 0.06527 & 0.09566 \\
$E_{{\rm B},\phi} / E_{n}$ & 0.04293 & 0.06724 \\
$E_{\rm th} / E_{n}$       & 0.5919  & 1.131   \\
Total $E / E_{n}$          & 2.243   & 2.906
\enddata
\end{deluxetable}

\begin{figure}
\epsscale{1.13}
\plotone{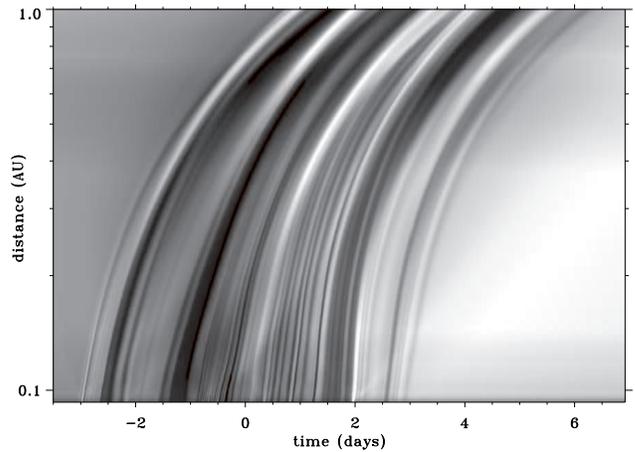}
\caption{Simulated time-distance variations of density fluctuations
from the high-diffusion MHD model.
The normalization of the horizontal time axis is the same as that
used in Figure \ref{fig13}.
Grayscale levels are defined independently at each radial distance
to be proportional to relative electron density variations.
\label{fig14}}
\end{figure}

We found it instructive to simulate global time-distance images
from the MHD models and compare them to similar images constructed
from observations of white-light heliospheric imagers.
For example, the Sun Earth Connection Coronal and Heliospheric
Investigation (SECCHI) on {\em STEREO} contains imagers that
probe electron density fluctuations at distances of order
15--300 $R_{\odot}$ \citep{Ey09}.
By stacking up radially oriented image slices taken over a range
of times, one can visualize the mutual interactions of fast
and slow streams, coronal mass ejections (CMEs), and shocks in
a clear way \citep[see, e.g.,][]{Dv09,Ro10,HD12}.
Figure \ref{fig14} shows the result, using the same time
coordinate as in Figure \ref{fig13} and using the mass density
$\rho$ as the plotted grayscale quantity.
At each height, the white and black limits of the grayscale map were
redefined based on the local minimum and maximum density in $\phi$.
Both the modeled and observed images show that fine-scale flux tube
variations in the innermost heliosphere become smeared out by
CIR-like stream interactions at larger distances.

Many of the results discussed above are independent of the adopted
value of the inner boundary distance.
As discussed above, we chose 20 $R_{\odot}$ as a standard baseline.
We wanted a value low enough that substantial stream interactions
would not have already occurred, but high enough that most of the
solar wind's acceleration as modeled by ZEPHYR would be included.
We also wanted to avoid the sub-Alfv\'{e}nic region in which
$u_{\phi}$ is undergoing a transition from rigid rotation
to angular momentum conservation \citep{WD67}.
In order to explore the sensitivity of the parameters at 1~AU to
the boundary distance, we ran two other low-diffusion models
with inner radii of 15 $R_{\odot}$ and 30 $R_{\odot}$.
Because we did not fine-tune the polytropic $\gamma$ to reproduce
the ZEPHYR code's wind acceleration profile at large distances,
we found that the mean wind speed $v_r$ at 1~AU does depend on
the choice of inner radius.
However, in all three models, the mean mass flux $\rho v_{r}$
and the mean radial field strength $B_r$ at 1~AU were nearly
invariant (i.e., they differed by less than 1\% from one another).
Also, the {\em relative} fluctuations at 1 AU, measured by the
ratios of standard deviations to mean values (for both $\rho$
and $v_r$) were nearly invariant across the three models.

\section{Statistics of Modeled MHD Fluctuations}
\label{sec:flucts}

Because the two-dimensional models described above do not include
any explicit time variability, the resulting ``fluctuations''
(as seen in, e.g., Figure \ref{fig13}) should not be expected to
resemble propagating MHD waves or turbulent eddies.
Nonetheless, it is useful to first examine them in the context of
linear MHD wave theory.
We determined the mean energy density components of the fluctuations
at 1 AU,
\begin{equation}
  E_{{\rm K},i} = \frac{\rho_{0} \langle v_{i}^{2} \rangle}{2}
  \,\, , \,\,\,\,
  E_{{\rm B},i} = \frac{\langle B_{i}^{2} \rangle}{8\pi}
  \,\, , \,\,\,\,
  E_{\rm th} = \frac{a_{0}^{2} \langle \rho^{2} \rangle}{2\rho_0}
  \,\, ,
  \label{eq:Evar}
\end{equation}
where quantities with subscript 0 refer to mean values taken over
the {80\arcdeg} of fully resolved longitudes, quantities given as
$\langle f^{2} \rangle$ refer to the variance of $f$ over the same
range of longitudes, and the subscript $i$ refers to either the
$r$ or $\phi$ vector components.
The three terms above refer to kinetic, magnetic, and thermal
fluctuations, respectively.
Alfv\'{e}nic fluctuations would be characterized by equipartition
between the kinetic and magnetic components perpendicular to the
background field.
Magnetosonic waves tend to have half of their fluctuation
energy in kinetic form and the other half divided between the
magnetic and thermal terms \citep[see Figure 2 of][]{CvB12}.
Pressure-balance structures (PBSs) are largely static features
that advect with the solar wind and carry along variations in gas
pressure and magnetic pressure that are in rough equipartition
between the $E_{\rm B}$ and $E_{\rm th}$ terms given above.

Table \ref{table01} gives the values of the fluctuation energy
density components for the two models discussed in
Section \ref{sec:cir}.
Each energy density has been normalized by dividing by a
representative 1~AU background magnetic energy density
$E_{n} = B_{n}^{2}/ 8\pi$, where
$B_{n} = 2 \times 10^{-5}$ G $=$ 2 nT.
This value of $B_n$ was chosen to be of the same order of magnitude
as the mean magnetic field strength at 1~AU in the models.
Table \ref{table01} shows that both models had significantly more
than half of their total energy in the form of kinetic fluctuations,
which is inconsistent with both linear MHD waves and PBSs.
Still, the models had an approximately magnetosonic balance
between the relative rms variations in magnetic field amplitude
($\langle B^{2} \rangle^{1/2} / B_{0}$) and density
($\langle \rho^{2} \rangle^{1/2} / \rho_{0}$), with both ratios
being roughly equal to 0.30 for the high-diffusion model and to
0.39 for the low-diffusion model.
This may be a natural consequence of the frozen-in MHD assumption,
since the spacing of field lines undergoes similar compressions
and rarefactions as the gas.

The total modeled energy density in fluctuations was between
2 and 3 times the background value $E_n$.
This can be compared with measured plasma fluctuations at 1 AU.
The total energy density of reported MHD turbulence is roughly
about 3--5~$E_n$ \citep[e.g.,][]{TM95,Ge95,BC05}, depending
on the type of solar wind stream.
This is in approximate agreement with the models, but we will see
below that there are key differences between the detailed
power spectra of the models and the observations.
We also found that larger-scale fluctuations in the ecliptic
plane contribute to a much higher energy density than do just the
small-scale turbulent motions.
For example, over the 16 days of OMNI data collected for
Figure \ref{fig12}(d), the {\em total} energy density computed
from all components of Equation (\ref{eq:Evar}) was found to
be 42.4~$E_n$.
Much of this (28.3~$E_n$) was in the kinetic energy variance
due to alternating fast and slow streams.
It is clear that the large-scale models presented here do
not come close to reproducing the full range of MHD variability
in the solar wind.

\begin{figure}
\epsscale{1.00}
\plotone{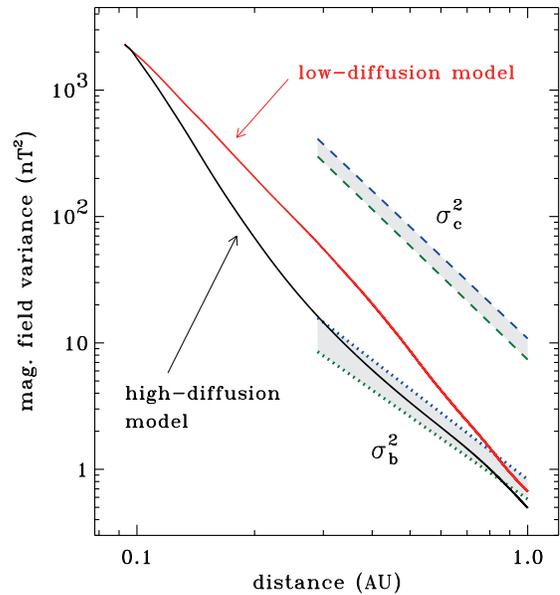}
\caption{Radial dependence of magnetic field variance in the
inner heliosphere.
For the low-diffusion model (red solid curve) and the
high-diffusion model (black solid curve), plotted values of
the component variance $\sigma_{c}^2$ are nearly identical to the 
magnitude variance $\sigma_{b}^2$ (not shown).
These are compared with {\em Helios} measurements of
$\sigma_{c}^2$ (dashed curves) and $\sigma_{b}^2$ (dotted curves)
for data averaged over 12 hour bins (blue upper curves) and
3 hour bins (green lower curves).
\label{fig15}}
\end{figure}

We can learn more about how the models may reproduce some aspects
of the actual solar wind by examining the radial evolution
of the magnetic field fluctuations.
Figure \ref{fig15} shows the radial dependence of the ``component
variance'' $\sigma_{c}^{2}$, which is the sum of the variances of
the individual vector components of ${\bf B}$.
For both the low-diffusion and high-diffusion models, this
quantity is nearly identical to (i.e., within 5\% of) the magnitude
variance $\sigma_{b}^{2}$, which is the variance of the scalar
field strength $| {\bf B} |$.
We do not plot $\sigma_{b}^{2}$ for the models since the curves
would be indistinguishable from the $\sigma_{c}^2$ curves.

Figure \ref{fig15} also compares the model results with measurements
made by {\em Helios} in the slow wind ($v \leq 500$ km~s$^{-1}$).
We show measured ranges for $\sigma_{c}^2$ and $\sigma_{b}^2$ given
by \citet{Mr78} for two different methods of binning the data.
In the real solar wind, $\sigma_{c}^{2} / \sigma_{b}^{2} \approx 10$,
which indicates that the directional variations in the vector
magnetic field greatly exceed the magnitude variations.
Although our model does not contain the small-scale MHD waves that
contribute to the observed $\sigma_{c}^2$, it does seem to reproduce
the large-scale magnitude variations that contribute to
$\sigma_{b}^2$.
Thus, we make the conjecture that the low-frequency variability of
magnetic field {\em magnitude} in the solar wind may be caused
by the longitudinal interaction and evolution of coronal flux tubes.

\begin{figure}
\epsscale{1.00}
\plotone{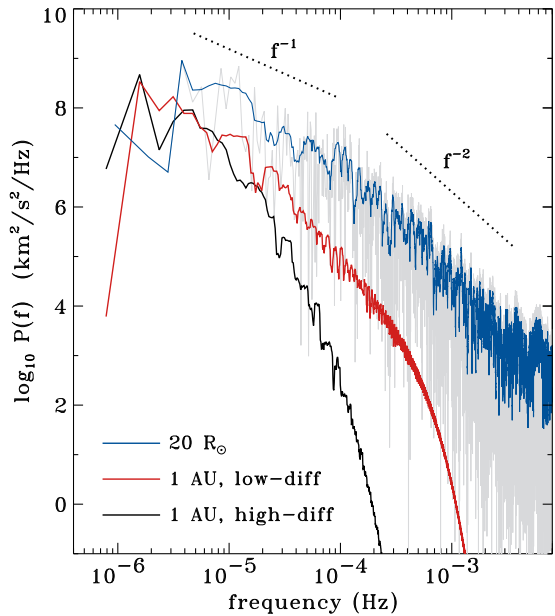}
\caption{Fourier power spectra of $v_r$ fluctuations in the
time-steady MHD models.
The raw FFT result for the inner boundary condition at
20 $R_{\odot}$ is shown (light gray curve) along with its
9-point boxcar smoothing average (blue curve).
Corresponding smoothed spectra at 1 AU are shown for the
low-diffusion model (red curve) and high-diffusion model
(black curve).
\label{fig16}}
\end{figure}

Additional information about the nature of the modeled fluctuations
can be extracted from their power spectra.
Converting the longitudinal coordinate $\phi$ into time using
Equation (\ref{eq:phitime}), we took fast Fourier transforms (FFTs)
of various MHD quantities over the {80\arcdeg} of fully resolved
longitude and plotted the power as a function of spacecraft-frame
frequency.
For example, Figure \ref{fig16} shows power spectra of the
fluctuations of radial velocity $v_r$ at both the inner boundary
radius of 20 $R_{\odot}$ and at 1~AU.
The curve at 20 $R_{\odot}$ is the same for the low-diffusion
and high-diffusion models because the same ZEPHYR boundary condition
was used for both.
Despite the high resolution of the models, the numerically determined
power spectra were extremely noisy.
Thus, for clarity we plot 9-point boxcar-smoothed curves for all
modeled spectra, and we show the full, unsmoothed spectrum for only
one case: the inner boundary spectrum at 20 $R_{\odot}$.
The smoothing does not alter the overall shape or magnitude of
the spectrum.

The velocity power spectrum at 20 $R_{\odot}$ is roughly
proportional to $f^{-1}$ for frequencies between about
$5 \times 10^{-6}$ and $10^{-4}$ Hz.
This is reminiscent of---but probably not causally related to---the
$f^{-1}$ spectrum seen for some quantities at frequencies
below $10^{-4}$ Hz in the solar wind at 1~AU \citep{Bv82,MG86,Ni09}.
The modeled spectrum steepens to roughly $f^{-2}$ at higher
frequencies, and the steepening continues as the spectrum evolves
outward in heliocentric distance.
The high-diffusion model steepens more than the low-diffusion
model.
This is expected since the former model undergoes more longitudinal
smearing of structure at small spatial scales (i.e., at high
spacecraft-frame frequencies).
The power at the very lowest frequencies ($f \lesssim
3 \times 10^{-6}$ Hz, i.e., timescales greater than a few days)
appears to grow with increasing heliocentric distance.
Although there is no formal ``inverse cascade'' in this model,
the transfer of energy from small to large scales can be the
result of CIR-like stream collisions that eventually produce
merged interaction regions \citep[MIRs; e.g.,][]{Bu03}.

\begin{figure}
\epsscale{1.00}
\plotone{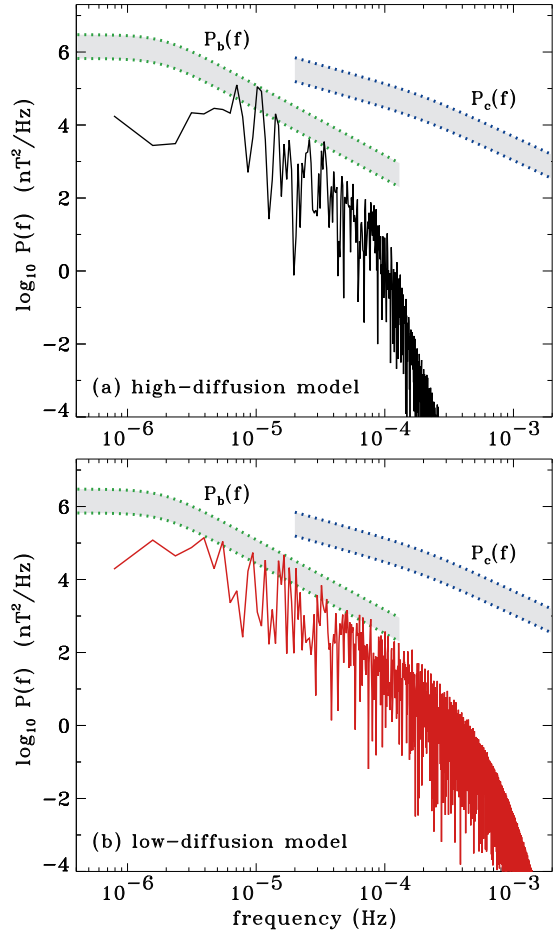}
\caption{Fourier power spectra of magnetic field fluctuations
in the high-diffusion (a) and low-diffusion (b) models at 1 AU.
Simulated power spectra (solid curves) are compared with
measured ranges of vector component fluctuations (blue dotted
curves) and scalar magnitude fluctuations (green dotted curves).
\label{fig17}}
\end{figure}

Figure \ref{fig17} shows the modeled magnetic fluctuation power
spectra at 1~AU without any boxcar smoothing.
For clarity, the results of the low-diffusion and high-diffusion
models are shown in separate panels.
The gray regions give the ranges of measured magnetic power
spectra for the slow solar wind in the ecliptic plane.
The spectrum labeled $P_{c}(f)$ gives the ``full'' MHD turbulence
spectrum that is analogous to the component variance $\sigma_{c}^2$
discussed above \citep[see, e.g.,][]{TM95,Ms02,BC05,Vb07}.
The spectrum labeled $P_{b}(f)$ is the spectrum of fluctuations
in the scalar field strength \citep{BM89}, and it is analogous to
the magnitude variance $\sigma_{b}^2$.
The agreement between the low-diffusion model spectrum and
the measured range of $P_{b}(f)$ is quite good.
This appears to reinforce our earlier suggestion that the
proposed model of longitudinally evolving coronal flux tubes can
explain the measured low-frequency fluctuations of magnetic field
magnitude in the solar wind.

It is also evident from Figure \ref{fig17} that the low-frequency
$f^{-1}$ part of the measured fluctuation spectrum $P_{c}(f)$
contains significantly more power than is found in the modeled
spectra at 1~AU.
The $f^{-1}$ fluctuations are often claimed to be
fossil remnants of coronal variations \citep[e.g.,][]{Rb10}.
However, we can tentatively conclude from these simulations that
if this is true, it is more likely for them to represent propagating
waves and not just the ``passive'' survival of coronal flux tubes.

\section{Discussion and Conclusions}
\label{sec:conc}

The aim of this paper was to begin exploring how MHD fluctuations
in interplanetary space are related to the well-known filamentary
flux-tube structure of the solar corona.
Starting with a time-steady potential-field model of the QS corona
allowed us to model the properties of a ``bundle'' of open field
lines with footpoint separations that extend down to scales smaller
than those of both granules and intergranular lanes.
This let us study magnetic field variations on spatial scales
that are still unresolvable with state-of-the-art MHD simulations
of the global corona and heliosphere.
After adjusting the field strengths found from the PFSS approximation,
we produced one-dimensional models of the coronal heating and solar
wind acceleration along these field lines.
These models assumed that the energization of the corona comes from
the dissipation of MHD turbulence, but in principle any model
of plasma heating could be applied.
We also used the one-dimensional results as lower boundary conditions
to a model of radial and longitudinal heliospheric evolution, and we
predicted the time-steady formation of CIR structure from the
coronal flux tubes.
The resulting fluctuations in magnetic field {\em magnitude} greatly
resemble corresponding fluctuations that have been measured in
the low-latitude slow solar wind.

One original goal of this paper was to study the origin of
low-frequency MHD fluctuations in the solar wind.
Power spectra obeying a relatively flat $f^{-1}$ frequency
dependence are often suggested to be the turbulent ``energy
containing range'' and of direct solar origin, but their cause
is still being debated \citep{MG86,Ve89,Mt07,Dm07,Dm11,Vd12b}.
There is also still disagreement over whether the
statistical properties of MHD discontinuities are due to flux
tubes of solar origin \citep{Bo10} or are the by-product of
an ongoing turbulent cascade \citep{Gr08}.
Our ecliptic-plane models were not set up to reproduce
inherently time-variable fluctuations, but they do help
put limits on the ability of small-scale coronal variations
to survive the effects of stream-stream CIR-type interactions.
They also act as a basal level of expected {\em background
fluctuations} to which must be added any propagating waves
or turbulent motions.
Interactions between waves and static structures have been
invoked as possible means of mode conversion and coronal heating
\citep[e.g.,][]{Go07,Ev12}, so knowing more about the background
variations is important to help constrain those theories.

As emphasized in Section \ref{sec:solis}, our time-steady
description of the coronal magnetic field was meant only to be a
snapshot that is representative of the high-resolution topology
that exists at any one time \citep[see also][]{JP06}.
The validity of this kind of model depends crucially on the scales
being resolved.
Because the smallest structures will have the shortest lifetimes,
it becomes increasingly necessary, when going to smaller scales,
to take into account the evolving nature of the magnetic carpet.
\citet{CvB10}, \citet{MvB12}, and others presented Monte Carlo
simulations that may be useful to ``connect'' to the solar wind
acceleration models of this paper. 
At the very least, such simulations can be used to provide
insight about how rapidly the field geometry evolves as a function
of height.\footnote{%
Above a certain height, the magnetic carpet's evolution may be
slow enough that a parcel of solar wind can accelerate up to that
height without its footpoints being appreciably deformed.
In that case, our ``snapshot'' method may be better justified.}
Of course, fully time-dependent and multidimensional techniques
will be needed to describe the self-consistent interactions
between flux tubes, waves, turbulent fluctuations, and fast/slow
streams in the heliosphere.

We also note that our assumption of a potential field for
extrapolating the measured photospheric magnetogram data into the
corona is likely to be inaccurate in many key ways.
Using an improved MHD treatment could naturally incorporate many
of the non-potential effects that we attempted to account for in
our $B_2$ enhancement term.
In addition, the PFSS technique is known to eliminate nearly all
of the {\em smallest-scale} longitudinal flux-tube structure in
the field at and above the source surface.
In our case, this resulted in an inability to test the alternate
flux-tube expansion formalism of \citet{Fu05} and \citet{Sz06},
who suggested that $B_{\rm ss} \propto B_{\rm base} / f_{\rm ss}$
may be a better quantity to correlate with the wind speed than
$f_{\rm ss}$ itself.

Despite the above warnings, we believe the high-resolution magnetic
field models constructed for this paper can be useful tools for
studying the statistical properties of spatial plasma fluctuations
that exist sufficiently near the Sun that other effects (e.g., stream
interactions or turbulence) do not smear them out.
These variations can be compared straightforwardly to existing
remote-sensing observations.
For example, we would like to know if these inter-flux-tube
variations contribute to the measured nonthermal widths of
coronal emission lines \citep[e.g.,][]{Gu10} or to the large
velocity variances obtained from radio scintillation
data \citep{HC05}.
Both of these are commonly interpreted as being caused by the
line-of-sight integration through a corona filled by
Alfv\'{e}n waves, but the effects of ``static'' flux tubes remain
to be disentangled.
We would also like to simulate the MHD fluctuations at the
intermediate distances to be probed by {\em Solar Probe Plus}
($r \gtrsim 0.05 \, \mbox{AU}$) and {\em Solar Orbiter}
($r \gtrsim 0.28 \, \mbox{AU}$).
This requires a more self-consistent calculation of $u_{\phi}$
to model the turnover from rigid corotation to angular momentum
conservation \citep[e.g.,][]{WD67,PP74}, and it may also require
including ``true'' heating and acceleration source terms into
the two-dimensional code described in Section \ref{sec:cir}.

\acknowledgments

The authors gratefully acknowledge Bill Matthaeus,
Spiro Antiochos, Zoran Miki\'{c}, Aaron Roberts, and the
anonymous referee for valuable discussions.
This work was supported by the National Aeronautics and Space
Administration (NASA) under grants {NNX\-09\-AB27G} and
{NNX\-10\-AC11G} to the Smithsonian Astrophysical Observatory.
The SOLIS data used in this paper are produced cooperatively
by NSF/NSO and NASA/LWS.
The OMNI solar wind data were obtained from the GSFC/SPDF OMNIWeb
interface,
and we thank Dave McComas and Ruth Skoug ({\em{ACE}}/SWEPAM)
and Chuck Smith and Norm Ness ({\em{ACE}}/MAG) for providing the
majority of the OMNI measurements that were used in this paper.

\appendix

\section{Potential Field Source Surface Model}
\label{appen:pfss}

In this paper we compute the magnetic field assuming the corona is
current-free (i.e., a potential field).
We use a spherical domain extending from the solar surface up to a
radial distance $R_{\rm ss}$, where the field is assumed to be radial.
Low resolution versions of such potential field source surface (PFSS)
models have been used extensively in the past
\citep[e.g.,][]{Sh69,AN69,WS90,AP00,Lu02,WS06}.
In the present work we need high spatial resolution in the low
corona to accurately trace the roots of the open field lines on
scales below $\sim$1 Mm in the photosphere.
Using the standard method, which is based on spherical harmonic
expansion, very large grids covering the entire Sun would be required.
However, high resolution is needed only in a limited region on the
solar surface.
Therefore, we use a different computational method in which the high
resolution grid extends only over some limited area.
The basic methodology is described in Appendix B of \citet{vanB2000},
and can be summarized as follows.
The magnetic field ${\bf B} ({\bf r})$ is written in terms of a
vector potential ${\bf A} ({\bf r})$, which is given by
\begin{equation}
  A_{r} \, = \, 0 \, ,
  \,\,\,\,\,\,\,
  A_{\theta} \, = \, - \frac{1}{r \sin \theta}
  \frac{\partial \Psi}{\partial \phi} \, ,
\end{equation}
\begin{equation}
  A_{\phi} \, = \, \frac{1}{r} \left[ \frac{\partial \Psi}
  {\partial \theta} + B_{0} \tan ( \theta / 2) \right] \, ,
\end{equation}
where $(r, \theta, \phi)$ are spherical coordinates, $B_0$ is the
monopole component of the imposed radial field, and
$\Psi (r, \theta, \phi)$ is a scalar function that must satisfy
the following partial differential equation: 
\begin{equation}
  \frac{\partial^2 \Psi} {\partial r^2} +
  \nabla_{\perp}^2 \Psi \, = \, 0 \,\, .
  \label{eq:Psi2}
\end{equation}
The gradient operator $\nabla_{\perp}$ operates only on the
coordinates perpendicular to the radial direction.
Thus, the potential field is given by
\begin{equation}
  {\bf B} ({\bf r}) \, = \, B_{0}
  \left( \frac{R_{\sun}}{r} \right)^{2} \hat{\bf r}
  - \nabla \left( \frac{\partial \Psi}{\partial r} \right)
  \,\, .
\end{equation}
At the longitudinal boundaries of the high-resolution domain,
$B_\phi$ either vanishes or is periodic in longitude;
at the latitudinal boundaries we require $B_\theta = 0$.
Due to these boundary conditions, the solution of
Equation (\ref{eq:Psi2}) can be written as a superposition of
discrete eigenmodes, but the eigenmodes are not spherical harmonics
and must be computed numerically.
The mode amplitudes are chosen such that ${\bf B} ({\bf r})$ matches
the imposed radial magnetic field at the solar surface
($r = R_{\sun}$) and satisfies
$\partial \Psi / \partial r = 0$ at the source surface
($r = R_{\rm ss}$).

Two important improvements have been made to the basic method.
First, we use a variable grid.
At certain heights in the corona the grid spacing is doubled in
all three directions, allowing us to cover the full height range
of the model with a relatively small number of grid points
\citep[for details see Appendix A of][]{Bobra2008}.
Second, a description of the global corona has been added
to improve the side boundary conditions on the high-resolution domain
\citep[see Section 4.1 of][]{Su2011}.
The high-resolution and global models are computed in such a way
that the magnetic field is continuous across the side boundaries
of the high-resolution domain.
Hence, field lines can be traced through the side boundaries without
any unphysical kinks in the field lines.

As described above, we used a SOLIS line-of-sight (LOS)
photospheric magnetogram taken on 2003 September 4 at 16:30 UT,
as well as the SOLIS synoptic map for the relevant Carrington
Rotation (CR 2007).
The LOS magnetogram was converted into a longitude-latitude
map of the radial field $B_{r} (R_{\sun}, \theta, \phi)$ in the
high resolution part of the grid.
This conversion assumes that the magnetic field on the Sun is
nearly radial, which is a good approximation for the Quiet Sun.
The high resolution grid has $1024 \times 512$ cells on
the photosphere, and the grid spacing is
$0.0018 \cos \lambda \, R_\sun$, where $\lambda$ is the latitude.
The source surface is located at $R_{\rm ss} = 2.46 \, R_{\sun}$.
The synoptic map from CR 2007 was used for constructing the global
part of the PFSS model.

\hspace*{0.01in}

\hspace*{0.01in}

\end{document}